\documentclass[12,floatfix]{article}
\usepackage{graphicx,color}
\usepackage{amsmath}

\usepackage[bookmarksnumbered,bookmarksopen,colorlinks,citecolor=blue,linkcolor=blue]{hyperref}

\renewcommand\a{\alpha}

\renewcommand\d{\delta}

\renewcommand\o{\omega}

\newcommand\m{\mu}

\newcommand\x{\xi}
\newcommand\p{\pi}
\newcommand\h{\theta}

\newcommand\f{\phi}

\renewcommand\L{\Lambda}

\renewcommand\O{\Omega}

\newcommand{\fig}[1]{Fig.~\ref{#1}}
\newcommand{\eq}[1]{Eq.~(\ref{#1})}
\newcommand{\eqs}[2]{Eqs.~(\ref{#1})-(\ref{#2})}

\newcommand\lb{\left(}
\newcommand\rb{\right)}
\newcommand\ls{\left[}
\newcommand\rs{\right]}
\newcommand\lc{\left\{}
\newcommand\rc{\right\}}

\newcommand\ra{\rightarrow}

\newcommand{\non}{\nonumber\\}
\newcommand\pt{\partial}

\newcommand{\bx}{{\mathbf x}}

\newcommand{\bp}{{\mathbf p}}
\newcommand{\bk}{{\mathbf k}}
\newcommand{\bq}{{\mathbf q}}

\newcommand{\bv}{{\bf v}}
\newcommand{\bl}{{\bf l}}

\renewcommand{\part}{{\rm part}}

\renewcommand{\vec}{\boldsymbol}
\newcommand{\be}{\begin{equation}}
\newcommand{\ee}{\end{equation}}
\newcommand{\bear}{\begin{eqnarray}}
\newcommand{\eear}{\end{eqnarray}}
\newcommand{\ba}{\begin{array}}
\newcommand{\ea}{\end{array}}

\begin{document}

\title{Glasma Evolution and Bose-Einstein Condensation with Elastic and Inelastic Collisions}

\author{Xu-Guang Huang${}^{(1,2)}$,  Jinfeng Liao${}^{(2,3)}$}

\maketitle

\begin{enumerate}
\item   Physics Department and Center for Particle Physics and Field Theory,
Fudan University, Shanghai 200433, China.
\item   Physics Department and Center for Exploration of Energy and Matter,
Indiana University, 2401 N Milo B. Sampson Lane, Bloomington, IN 47408, USA.
 \item RIKEN BNL Research Center, Bldg. 510A, Brookhaven National Laboratory,
   Upton, NY 11973, USA.
\end{enumerate}

\begin{abstract}
In this paper we investigate the role of inelastic collisions in the kinetic evolution  of a highly overpopulated gluon system starting from Glasma-type initial condition. Using the Gunion-Bertsch formula we derive the inelastic collision kernel under the collinear and small angle approximations. With both numerics and analytic analysis, we show that the inelastic process has two effects: globally changing (mostly reducing) the total particle number, while locally at small momentum regime always filling up the infrared modes extremely quickly. This latter effect is found to significantly speed up the emergence of a local thermal distribution in the infrared regime with vanishing local ``chemical potential''  and thus catalyze the onset of dynamical Bose-Einstein Condensation to occur faster (as compared with the purely elastic case) in the overpopulated Glasma.
\end{abstract}


\newpage

\section{Introduction}

Thermalization of the quark-gluon plasma is one of the most challenging problems in current heavy ion physics. See e.g. Ref.~\cite{Berges:2012ks,Arnold:2007pg,Huang:2014iwa} for recent reviews. Starting with  two colliding nuclei in a form of color glass condensate with high gluon occupation $f\sim1/\alpha_{\rm s}$ below saturation scale $Q_{\rm s}$ \cite{McLerran:1993ni,Blaizot:1987nc,Iancu:2003xm} and following the initial impact,  a subsequent strong field evolution stage (likely with instabilities \cite{Romatschke:2005pm}) till about the time $1/Q_{\rm s}$ is then succeeded by a far-from-equilibrium gluon-dominant matter, the Glasma \cite{Lappi:2006fp}. The evolution of this Glasma stage toward a quark-gluon plasma (QGP) that is close to local equilibratium and exhibits viscous-hydrodynamic behavior, is  indicated by phenomenology to be reached on the order of a fermi over c time (see e.g. \cite{Heinz:2009xj}). Precisely how this occurs remains to be fully understood. Describing the pre-equilibrium evolution with kinetic equations is a very useful approach, based on which the so-called ``bottom-up'' thermalization scenario was developed \cite{Mueller:1999fp,Baier:2000sb,Mueller:2005un}. There is however the complication of
instability driven by anisotropy that may change this picture (see e.g. \cite{Arnold:2003rq,Mrowczynski:1993qm,Romatschke:2003ms,Rebhan:2005re}). There are also other kinetic-based approaches, see e.g. \cite{Xu:2004mz}.

More recently an alternative thermalization scenario, based on crucial role of high initial overpopulation in the Glasma and kinetic evolution dominated by elastic collisions, has been proposed in \cite{Blaizot:2011xf,BLM}. In this scenario, while the initial scale $Q_s$ is large compared with $\L_{QCD}$ and thus the coupling $\a_s$ is small, the high occupation $f\sim 1/\a_s$ elevates the elastic scattering rate to be of the order $\hat{O}(1)$ rather than the usual $\hat{O}(\a_s^2)$, and the Glasma is essentially an emergent strongly interacting matter with weak coupling albeit large aggregate of constituents. Two important scales are introduced to characterize the distribution, the hard cut-off scale $\L$ beyond which $f\ll 1$ and the soft high-occupation scale $\L_s$ below which $f\sim 1/\a_s$. While the initial Glasma has the two scales overlapping $\L\sim \L_s \sim Q_s$, during the course of thermalization the two scales are separated eventually toward $\L_s \sim \a_s \L$ upon thermalization.
One particularly nontrivial observation in the elastic-dominant picture is that the high initial overpopulation $n / \epsilon^{3/4} \sim 1/a_s^{1/4}\gg1$ and the conservation of both energy and particle number will necessarily require the formation of a Bose-Einstein condensate that absorbs the excess gluons. This has been explicitly shown to occur by numerically solving the elastic kinetic equation derived under small angle approximation \cite{BLM}. There have been intensive discussions related to this picture from a variety of approaches, see e.g. \cite{Kurkela:2011ti,Kurkela:2011ub,Epelbaum:2011pc,Gelis:2011xw,Berges:2012us,Berges:2011sb,Berges:2012ev,Berges:2012mc,Kurkela:2012hp,Schlichting:2012es,York:2014wja,Berges:2013eia,Gelis:2013rba,Attems:2012js,Chiu:2012ij,Blaizot:2012qd,Liao:2012qk,Ruggieri:2013ova}.
Strong evidences for the formation of such a Bose-Einstein condensate have been reported for similar thermalization problem in the classical-statistical lattice simulation of scalar field theory \cite{Epelbaum:2011pc,Gelis:2011xw,Berges:2012us}. The case for non-Abelian gauge theory is more complicated and still under investigation \cite{Berges:2011sb,Berges:2012ev,Berges:2012mc,Kurkela:2012hp,Schlichting:2012es,York:2014wja}.

One important question that has not been addressed in the above scenario is the role of inelastic processes. This issue could indeed be critical for at least two reasons (see discussions in e.g. \cite{Blaizot:2011xf,BLM,Kurkela:2011ti,Kurkela:2011ub,Blaizot:2012qd,Liao:2012qk}).  First of all the inelastic processes will spoil the particle number conservation, and one might naively argue that the excessive gluons in the overpopulated Glasma could simply be eliminated by very fast inelastic collisions. Secondly, to make it even worse, the inelastic processes are parametrically at the same order as the elastic processes (as opposed to naive power counting), so there appears no apparent dominance of the elastic  over the inelastic and one may indeed worry that the inelastic could efficiently reduce total particle number. In such a situation, an explicit evaluation including both elastic and inelastic collisions becomes mandatory to clarify what will happen after including both types of collisions. To be precise, once the inelastic processes are included, one does not expect any condensation in the ultimate thermal equilibrium because with long enough time the inelastic processes will always remove any excessive particles. The interesting question, instead, is what changes the inelastic collisions bring to the dynamical evolution of the system. In particular, it is found ~\cite{BLM} that with purely elastic scatterings the overpopulated system is driven toward a dynamical onset of condensation in a finite time via critical scaling behavior in the infrared regime. It is extremely interesting to know, upon including the inelastic processes, how such dynamical evolution may be modified and whether the transient off-equilibrium condensation would still occur or not.

In this paper, we aim to address this important question by studying the kinetic evolution of a highly overpopulated system starting from Glasma-type initial condition with both $2\leftrightarrow 2$ and $2\leftrightarrow 3$ scatterings. In Section 2  we will derive the inelastic collision kernel under the collinear and small angle approximations using the Gunion-Bertsch formula for the $2\leftrightarrow3$ matrix element. In Section 3 we will use numerical solutions as well as analytic analysis to understand the role of  the inelastic process for both the global particle number change and the local behavior at small momentum region. Finally we will conclude in Section 4. As a first step toward understanding the inelastic contributions and for simplicity and unambiguity, we will focus on the static box case with isotropic distribution in this work and leave the study of expanding case for future work.

It may be noted that the kinetic theory framework is best suited for studying well-defined quasi-particle excitations at typical scales in a physical system. Pushing the use of this approach into the deep infrared regime may bear theoretical issues that are not easily clarified. One however may notice that the kinetic description has been widely adopted for studying the Bose-Einstein Condensation phenomena across a wide range of physical systems, e.g. for cosmological scalars~\cite{Semikoz:1994zp,Semikoz:1995rd}, for general Bose gases with varied interactions~\cite{gas_1,gas_2,gas_3,gas_4}, for trapped atomic gases~\cite{atomic_bec}, as well as for polaritons in condensed matter systems~\cite{polariton,polariton_2}. In particular the kinetic equations are shown in the above literature to be a very useful tool in understanding the BEC onset which is a non-equilibrium process. Additionally, it shall be emphasized that the mathematical properties of kinetic equations are of their own interests. The kinetic equations have well defined fixed point solutions  (which may contain a condensate in the overpopulated case), and the detailed evolution of the distribution function toward such solutions is highly nontrivial and interesting to know. We therefore believe the present kinetic theory study is a plausible approach for gaining useful insights about the evolution and possible onset of Bose-Einstein Condensation in the overpopulated glasma.

\section{Kinetic Evolution with Elastic and Inelastic Collisions}

In this section we will derive the kinetic evolution equation with both elastic and inelastic collisions. The kinetic equation deals with the gluon distribution function defined as
\begin{eqnarray}
f(t,\bx,\bp)\equiv\frac{(2\p)^3}{N_g}\frac{dN}{d^3\bx d^3\bp},
\end{eqnarray}
where $N_g=2(N_c^2-1)$ denotes the spin and color degeneracy factor.
The Boltzmann equation for $f(t,\bx,\bp)$ reads
\begin{eqnarray}
{\cal D}_t f_p={\cal C}_{2\leftrightarrow 2}[f_p]+{\cal C}_{2\leftrightarrow 3}[f_p],
\end{eqnarray}
where we denote $f(t,\bx,\bp)$ by $f_p$ and
\begin{eqnarray}
{\cal D}_t \equiv\frac{p^\m}{E_p}\pt_\m=\pt_t+\bv_p\cdot\nabla_\bx
\end{eqnarray}
with $\bv_p\equiv\bp/E_p$ and $E_p=|\bp|$. For later convenience, we also introduce the following notations:
\begin{eqnarray}
g_p \equiv 1 + f_p \,\, , \,\, h_p \equiv f_p \, g_p = f_p(1+f_p) \,\, .
\end{eqnarray}
In what follows we will separately discuss the elastic term ${\cal C}_{2\leftrightarrow 2}$ and the inelastic term ${\cal C}_{2\leftrightarrow 3}$.

\subsection{The $2\leftrightarrow 2$ process}

The collision kernel from the $2 \leftrightarrow 2$ process with full nonlinearity has been studied in \cite{Blaizot:2011xf,BLM}. Here we only briefly summarize the main results. We have the $2\leftrightarrow 2$ collision kernel given by
\begin{eqnarray}
{\cal C}_{2\leftrightarrow 2}[f_p]&=&\frac{1}{N_g}\frac{1}{2}\int_{123}\frac{1}{2E_p}|M_{12\leftrightarrow3p}|^2
(2\p)^4\d^4(p_1+p_2-p_3-p)\non&&\times[(1+f_p)(1+f_3)f_1f_2-f_pf_3(1+f_1)(1+f_2)], \label{C22_full}
\end{eqnarray}
where
\begin{eqnarray}
\int_i\equiv\int\frac{d^3\bp_i}{(2\p)^32E_i},
\end{eqnarray}
and
\begin{eqnarray}
|M_{12\leftrightarrow3p}|^2=8g^4N_c^2N_g\lb3-\frac{tu}{s^2}-\frac{su}{t^2}-\frac{ts}{u^2}\rb \label{M22}
\end{eqnarray}
is the (squared) $2\leftrightarrow2$ collision matrix element with $s=(p+p_3)^2, t=(p-p_1)^2, u=(p-p_2)^2$ being the usual Mandelstam variables.
The pre-factor $1/2$ in \eq{C22_full} is a symmetry factor counterweighing the permutation of $1$ and $2$ while the pre-factor $1/N_g$ cancels the summation over the spin and color of gluon ``$p$" in the matrix element (\ref{M22}).

The dominant contribution of $2\leftrightarrow 2$ scattering in (\ref{M22}) comes from very small exchange momentum in $t\to 0$ or $u\to 0$ kinematic regimes, for which the incoming momenta only gets ``deflected'' by very small angle. If one uses this small angle approximation, then a rather neat kernel can be derived \cite{BLM} :
\begin{eqnarray}
{\cal C}_{2\leftrightarrow 2}=\xi \L_s^2\L\frac{1}{p^2}\pt_p\lc p^2\ls\frac{\pt f_p}{\pt p}+\frac{\a_s}{\L_s}f_p(1+f_p)\rs\rc, \label{C22_sa}
\end{eqnarray}
with $\xi =(2N_c^2/\p)\int dq/q$ coming from the leading-log contribution. The hard scale $\Lambda$ and soft scale $\Lambda_s$ in the above are defined via global integrals:
\begin{eqnarray}
\L\L_s^2 / \a_s^2 &=&\int_0^\infty dpp^2 f_p(1+f_p) \equiv I_a \, ,\\
\L\L_s / \a_s   &=&  \int_0^\infty dpp^2 (2f_p/p) \equiv I_b \, .
\end{eqnarray}
For later convenience we also introduce the Debye scale  defined as \cite{Braaten:1989mz,Blaizot:2001nr}
\begin{eqnarray}
m_D^2  = -\a_s \int_0^\infty dpp^2  \partial f_p / \partial p =  \L\L_s.
\end{eqnarray}
It is interesting to notice that in a weakly coupled thermal QGP one has the well-defined separation of scales, $\L\sim T$, $m_D\sim gT$, $\L_s\sim g^2T$. The matter becomes strongly interacting when the scales ``collapse'' together. One way for that to happen is to have the system become really strongly coupled $g\to 1$ which likely will be accompanied by change of underlying degrees of freedom \cite{Liao:2006ry}. The other possibility, as in the case of Glasma, is when the system is highly off-equilibrium and overpopulated $f\sim 1/g^2$ --- in this case all the scales also become of the same order $\L \sim m_D \sim \L_s \sim Q_s $ and make the system emerge as a strongly interacting matter.

Clearly, both the full form ${\cal C}_{2\leftrightarrow 2}$ in (\ref{C22_full}) and the small angle approximation form in (\ref{C22_sa}) conserve the energy as well as particle number, as they should. In addition the Bose-Einstein distribution $f_{BE} = 1/ [e^{(p-\mu)/T}-1]$ with any $T$ and $\mu$ (in correspondence to the two conserved quantities) is the fixed point solution that makes both (\ref{C22_full}) and (\ref{C22_sa}) vanish. As a cautionary remark, one may notice that the small angle approximation may become questionable in low momentum regime and medium screening effects may also require improvements of the treatment here. Our main purpose though, is to understand the robust features of the dynamical onset process which may be not that sensitive to the details of such approximations. In the elastic scattering case, two very recent studies~\cite{Scardina:2014gxa,Xu:2014ega} have both studied the kinetic evolutions without the small angle approximations and have both confirmed the findings made in \cite{BLM} with small angle approximations. It is therefore conceivable that, keeping such caveats in mind, one can still learn useful lessons about the onset dynamics in the small angle approximations.

\subsection{The $2\leftrightarrow 3$ process}\label{transport}

\begin{figure}
\begin{center}
\includegraphics[width=5.5cm]{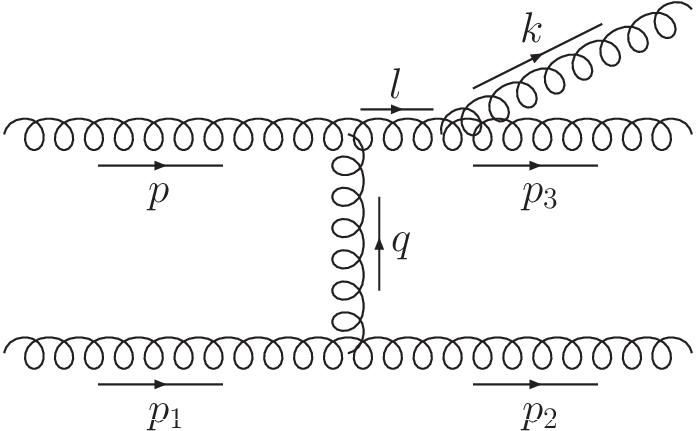}\;\;\;\;\;\;\;\;
\includegraphics[width=5.5cm]{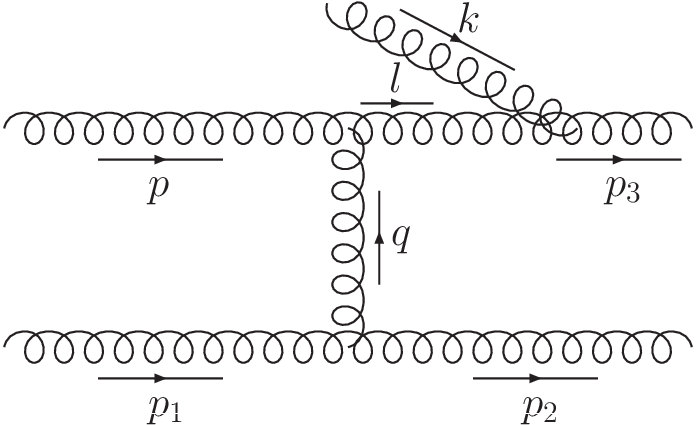}
\caption{(Left) A typical Feynman diagram contributing to ${\cal C}_{2 \leftrightarrow 3}^a$: $p+1 \leftrightarrow 2+3+k$.
(Right) A typical Feynman diagram contributing to ${\cal C}_{2 \leftrightarrow 3}^b$: $1+k+p \leftrightarrow 2+3$.}
\label{f2}
\end{center}
\vspace{-0.7cm}
\end{figure}

We now turn to the collision kernel from the $2\leftrightarrow 3$ process as depicted in Fig.\ref{f2}.
We denote the particle we are watching with momentum $p$, the softest external momentum with $k$, the exchange internal four-momentum with $q^\mu$, and then the rest external momenta with $p_{1,2,3}$.  The $2\leftrightarrow 3$ collision kernel can then be split into two pieces in which the particle $p$ is on the two-particle side or three-particle side respectively (see Fig.\ref{f2}):
\begin{eqnarray}
{\cal C}_{2\leftrightarrow 3}[f_p]&=&{\cal C}^a_{2 \leftrightarrow 3}+{\cal C}_{2 \leftrightarrow 3}^b,\\
{\cal C}_{2\leftrightarrow 3}^a&=&\frac{1}{N_g}\frac{1}{6}\int_{123k}\frac{1}{2E_p}|M_{1p  \leftrightarrow 23k}|^2
(2\p)^4\d^4(p+p_1-p_2-p_3-k)\non&&\times[(1+f_p)(1+f_1)f_2f_3f_k-f_pf_1(1+f_2)(1+f_3)(1+f_k)],\non
{\cal C}_{2 \leftrightarrow 3}^b&=&\frac{1}{N_g}\frac{1}{4}\int_{123k}\frac{1}{2E_p}|M_{23 \leftrightarrow 1kp}|^2
(2\p)^4\d^4(p+p_1+k-p_2-p_3)\non&&\times[(1+f_p)(1+f_1)(1+f_k)f_3f_2-f_pf_1f_k(1+f_3)(1+f_2)],
\end{eqnarray}
where the gluon labeled by $k$ will be treated as the soft emitted or absorbed gluon.
The factor $1/6$ counteracts
the $6$ equivalent permutations in $23k$ in process $1+p \leftrightarrow 2+3+k$ (see \fig{f2} left panel)
and the factor $1/4$ counteracts
the $4$ equivalent permutations in $1k$ and $23$ in process
$2+3 \leftrightarrow 1+k+p$ (see \fig{f2} right panel). We note that the graphs in \fig{f2} are used to
make the kinematics clear and it does not mean that only these two diagrams contribute: there are actually
25 different diagrams for ${\cal C}^a_{2 \leftrightarrow 3}$ and 25 diagrams for ${\cal C}^b_{2 \leftrightarrow 3}$. So the full matrix element $|M_{1p \leftrightarrow  23k}|^2$ is obtained by calculating 25 Feynman diagrams and it contains 6 equivalent kinematic setups in accordance with 6 permutations in $23k$ (see Appendix \ref{amplitude}).
We can then choose the kinematic setup corresponding to the \fig{f2} left panel, and multiply
6 to account other 5 kinematic setups. Similarly, we can fix the kinematics for $|M_{23 \leftrightarrow 1kp}|^2$ as in the
\fig{f2} right panel, and multiply 4 to get ${\cal C}^b_{2 \leftrightarrow  3}$. Thus we obtain
\begin{eqnarray}
{\cal C}_{2 \leftrightarrow 3}^a&=&\frac{1}{N_g}\int_{123k}\frac{1}{2E_p}|M^a_{1p\leftrightarrow23k}|^2
(2\p)^4\d^4(p+p_1-p_2-p_3-k)\non&&\times[(1+f_p)(1+f_1)f_2f_3f_k-f_pf_1(1+f_2)(1+f_3)(1+f_k)],\non
{\cal C}_{2 \leftrightarrow 3}^b&=&\frac{1}{N_g}\int_{123k}\frac{1}{2E_p}|M^b_{23\leftrightarrow1kp}|^2
(2\p)^4\d^4(p+p_1+k-p_2-p_3)\non&&\times[(1+f_p)(1+f_1)(1+f_k)f_2f_3-f_pf_1f_k(1+f_2)(1+f_3)],
\end{eqnarray}
where $|M^{a,b}|^2$ are the matrix element with the kinematics fixed according to \fig{f2}.
While the exact $2\leftrightarrow 3$ matrix element is known~\cite{Berends:1981rb}, it is hard to be directly used in a kinetic approach. Following many previous studies involving this process \cite{Gunion:1981qs,Arnold:2000dr,Xu:2007jv,Chen:2009sm}, we will use the so-called Gunion-Bertsch formula which is the collinear approximation and small angle approximation form of the exact matrix element and has been shown to give the dominant contribution in many cases.
Leaving the technical details to the Appendix \ref{amplitude}, we here quote the Gunion-Bertsch matrix element which is at the leading order in soft $q$ and $k$ expansion:
\begin{eqnarray}
|M^a_{1p \leftrightarrow 23k}|^2&=&64g^6N_c^3N_g\frac{(p\cdot p_1)^3}{q^2(q-k)^2(p\cdot k)(p_1\cdot k)},\non
|M^b_{23 \leftrightarrow 1kp}|^2&=&64g^6N_c^3N_g\frac{(p_2\cdot p_3)^3}{q^2(q+k)^2
(p_2\cdot k)(p_3\cdot k)}.
\end{eqnarray}

Like in the $2\leftrightarrow  2$ case, the matrix elements are dominated by the regime of very soft $q$ and $k$. We can thus further simplify the collision kernel using similar small angle approximation as in the elastic case. There is though additional subtlety as now there are two soft scales. In fact, as shown by the detailed analysis in Appendix \ref{colli}, the whole collision kernel can be separated into two pieces corresponding to contributions from different kinetic domains. (Note that both graphs in \fig{f2} contribute to each of these domains.) In the domain with $k$ being the softest scale, i.e. the ultra-soft emission and absorption, the $2\leftrightarrow 3$ essentially reduces to an effective $2 \leftrightarrow 2$ scattering with a collinear splitting/merging, and the resulting contribution to the collision kernel becomes
\begin{eqnarray}
{\cal C}^>_{2 \leftrightarrow  3}&\approx&\frac{1}{N_g}\int_{12l}\frac{1}{2E_p}|M_{1p\leftrightarrow2l}|^2 (2\p)^4\d^4(p+p_1-p_2-l) \nonumber \\
&& \times \left ( g_pg_1f_2f_l-f_pf_1g_2g_l \right) \, {\cal D}(|q=p_1-p_2|),
\end{eqnarray}
where the momentum labels $l$ and $q$ are as shown in \fig{f2}, and we also introduced the term ${\cal D}(|q=p_1-p_2|)$ arising from the splitting function integration. Its explicit form is given by \eq{defD} in Appendix \ref{colli}. For the above form, one can further simplify the effective $2\leftrightarrow  2$ part using small angle approximation as done in the elastic case. All the details are presented in Appendix \ref{colli}.
On the other hand in the domain with $q$ being the softest scale, the $2\leftrightarrow  3$ process effectively becomes a nearly collinear $1\leftrightarrow  2$ emission/combining process preceded by a small angle $2\leftrightarrow  2$ scattering that brings one incoming particle slightly off-shell. This part contributes the following to the collision kernel:
\begin{eqnarray}
{\cal C}^<_{2 \leftrightarrow 3}
&=& \frac{3g^6N_c^3}{16\p^5}
\int_0^\infty dp_1 p_1^2h_1\int_{-1}^1\frac{dx}{1-x}\int_0^\infty \frac{dq}{q^3} \nonumber \\
&&\times
\int_{0}^{z_c} \frac{dz}{z}\bigg\{
\ls g_pf_{(1-z)p}f_{zp}-f_pg_{(1-z)p}g_{zp} \rs \nonumber \\
&&\qquad +\frac{1}{(1-z)^4} \ls g_pg_{zp/(1-z)}f_{p/(1-z)}-f_pf_{zp/(1-z)}g_{p/(1-z)})\rs \bigg\}.
\end{eqnarray}
Note that in the $z$-integration we introduce an upper cut $z_c$: physically this is because that $k$ is the softest external momentum so the $k<p$ condition would require $z_c<1/2$.
In both kernels above, a number of infrared divergences appear. We will use the Debye scale $m_D$ as the infrared cut, e.g. $\int dq /q^3 \approx 1/m_D^2 = 1/(\L \L_s)$. We will also treat all leading logs as order one constant. Again more detailed discussions are included in the Appendix \ref{colli}.
Lastly, one can show that the inelastic kernel conserves energy while not particle number, and the fixed point solution (i.e. the equilibrium distribution) is the Bose-Einstein distribution {\it without} chemical potential, $f_{BE} = 1/(e^{p/T}-1)$.

It should be mentioned that for the inelastic processes, the inclusion of the so-called Landau-Pomeranchuk-Migdal (LPM) effect may bear important consequence (see e.g. \cite{Baier:2000sb,York:2014wja}). We though emphasize that in both cases (with or without LPM effect), the elastic and inelastic processes are parametrically at the same order and the final fixed point is the same Bose-Einstein distribution with zero chemical potential. This latter feature indicates that the inelastic processes will always tend to ``fill up'' the infrared modes even though the rates may differ in the cases with or without LPM effect. The inelastic kernel we have derived above, contains the most essential features of number-changing processes (as compared with the elastic), namely the non-conservation of particle number and the proper fixed point solution without chemical potential. It is therefore plausible that our study with the above inelastic kernel would capture the important qualitative influences of number-changing processes on the dynamical evolution before the BEC onset which is the main purpose of the present paper.

\subsection {The final kinetic equation}
\label{kine}
Finally we combine the ${\cal C}_{2 \leftrightarrow 2}$ and ${\cal C}_{2 \leftrightarrow3}$ kernels, and the final kinetic equation under small angle approximation
and collinear approximation reads
\begin{eqnarray} \label{kernel_all}
{\cal D}_t f_p={\cal C}^{\rm eff}_{2 \leftrightarrow 2}[f_p]+{\cal C}^{\rm eff}_{1 \leftrightarrow 2}[f_p],
\end{eqnarray}
where
\begin{eqnarray}
{\cal C}^{\rm eff}_{2 \leftrightarrow 2}={\cal C}_{2 \leftrightarrow 2}+{\cal C}^>_{2 \leftrightarrow 3}, \quad {\cal C}^{\rm eff}_{1 \leftrightarrow 2}={\cal C}^<_{2 \leftrightarrow 3}.
\end{eqnarray}
The expression for ${\cal C}^{\rm eff}_{2\ra2}$ is
\begin{eqnarray} \label{kernel_22}
{\cal C}^{\rm eff}_{2 \leftrightarrow 2}&=&\xi   \a_s^2 \, (1+D) \, I_a  \frac{1}{p^2}\pt_p\lc p^2\ls\frac{\pt f_p}{\pt p}+\frac{I_b}{I_a}f_p\rs\rc,
\end{eqnarray}
with $D\sim\hat{O}(1)$ parameterizing the contribution from ${\cal C}^>_{2 \leftrightarrow 3}$ to effective $2\leftrightarrow 2$ kernel. It is related to function ${\cal D}(|q|)$ defined in \eq{defD} by $D\sim {\cal D}(m_D)$.
 The expression for ${\cal C}^{\rm eff}_{1 \leftrightarrow 2}$ is
\begin{eqnarray} \label{kernel_12}
{\cal C}^{\rm eff}_{1 \leftrightarrow 2}&=&
 \x \, \a_s^2 \, R\, \frac{I_a}{I_b}
\bigg\{\int_{0}^{z_c} \frac{dz}{z}
\ls g_pf_{(1-z)p}f_{zp}-f_pg_{(1-z)p}g_{zp}
\rs\non&&+\int_{0}^{z_c} \frac{dz}{(1-z)^4z}
\ls g_pg_{zp/(1-z)}f_{p/(1-z)}-f_pf_{zp/(1-z)}g_{p/(1-z)})\rs\bigg\},
\end{eqnarray}
where the constant $R\sim\hat{O}(1)$ parameterizes the relative ratio of the order one constants between the elastic and inelastic kernels and  the cut $z_c$ in $z$-integration should be small to be consistent with the kinematics $k < p$. The expression for $R$ is given by \eq{defR} in Appendix \ref{colli}.

A few comments are in order here: \\
1) The kernel (\ref{kernel_22}) conserves energy and particle number with fixed point solution $f_{BE}=1/[e^{(p-\mu)/T}-1]$, while the kernel (\ref{kernel_12}) only conserves energy  with fixed point solution $f_{BE}=1/[e^{p/T}-1]$, so the total kernel (\ref{kernel_all}) conserves only energy and the equilibrium solution should be $f_{BE}=1/[e^{p/T}-1]$ without any chemical potential which is different from the pure elastic case; \\
2) In the nearly equilibrium case with $f\sim \hat{O}(1)$, $I_a \sim T^3$ and $I_b \sim T^2$, the elastic collision rate scales as $\sim \a_s^2 T$ and the inelastic rate scales also as $\sim \a_s^2 T$ so they are at parametrically the same order; \\
3) In the Glasma-like overpopulated case with $f\sim \hat{O}(1/\a_s)$, $I_a \sim Q_s^3/\a_s^2$ and $I_b \sim Q_s^2/\a_s$, the elastic collision rate scales as $\sim Q_s $ and the inelastic rate scales also as $\sim Q_s$ so again they are parametrically at the same order.\\
We therefore see that the effect of the inelastic collision is parametrically as important as the elastic one both near and far from equilibrium, and  including the inelastic collision qualitatively changes the ultimate equilibrium solution. It is clear that even starting from highly overpopulated initial condition, eventually there will be no chemical potential nor any condensate in the final thermal distribution with the presence of inelastic collision. However, the very important question that has not been understood, is how the inelastic collision will affect the transient dynamical off-equilibrium condensation driven by the pure elastic evolution starting from initial high overpopulation. Will the system still reach the onset of such condensation? Will the inelastic collision speed up, delay, or completely eliminate such an onset? We will address these questions by numerically solving the above kinetic equations.

Before turning to the numerical study, let us emphasize that the kinetic equations derived above are applicable only for describing the system till any moment before the actual onset of the BEC which is signaled by the emergence of an infrared singularity in the distribution. As is well known in the literature~\cite{Semikoz:1994zp,Semikoz:1995rd,gas_1,gas_2,gas_3}, kinetic theory breaks down at the onset point. After the formation of BEC, a modified kinetic theory framework is needed by explicitly introducing a condensate component. The growth of the condensate and the further evolution of the distribution should be described by a different set of kinetic equations that couple the condensate and particles together. In this paper we focus on understanding the evolution process from overpopulated initial conditions toward the onset of BEC for which our derived kinetic equations are suitable.

\section{Numerical Study of the Kinetic Evolution}

In this section, we numerically solve the kinetic equation (\ref{kernel_all}), starting with a Glasma-type initial condition as follows:
\begin{eqnarray} \label{initial}
f(p,t=0) =  f_0\, \theta(Q_s - p)
\end{eqnarray}
with $Q_s$ the saturation scale. We use $Q_s$ as unit for all momenta/enegy etc and use $1/Q_s$ as unit for time. As studied in \cite{BLM}, with a given initial occupation $f_0$, the overpopulation parameter is $n\epsilon^{-3/4} =  f_0^{1/4}\, 2^{5/4}/3\pi^{1/2}$ and when $f_0>f_0^c\approx 0.154$, the system is overpopulated as compared with the Stefan-Boltzmann limit and the system will reach onset of condensation when there is only elastic collision. For simplicity we fix the initial occupation $f_0=1$ which is in the overpopulated regime.
Note that the constant $\xi \a_s^2$ can be absorbed by a redefinition of time variable, $t \to (\xi \a_s^2) t$ which we will use from now on. We therefore are left with three parameters, $R$, $D$, and $z_c$. The inelastic contribution will increase with increasing $R$ and $z_c$ while the elastic will increase with increasing $D$. We will study the effect of inelastic collision by comparison with the purely elastic case ($R=0$ versus $R\ne 0$) and by varying the strength of the inelastic kernel.

\subsection{Thermalization in the purely inelastic case}
Let us first study the kinetic equation with only the inelastic kernel (\ref{kernel_12}). Although this is not a realistic modeling of the Glasma system, it is a very interesting problem on its own and it also serves as a very useful check of whether the derived inelastic kernel produces the physically expected dynamics. Furthermore it is a useful benchmark for a contrast with the evolution driven by both elastic and inelastic processes. We will choose $R=1$ without loss of generality because $R$ can be absorbed by redefining the time variable, $t\to R t$. The kinetic equation is then solved numerically with only the inelastic kernel and with overpopulated initial condition $f_0=1$.

\begin{figure}
\begin{center}
\includegraphics[width=5.55cm]{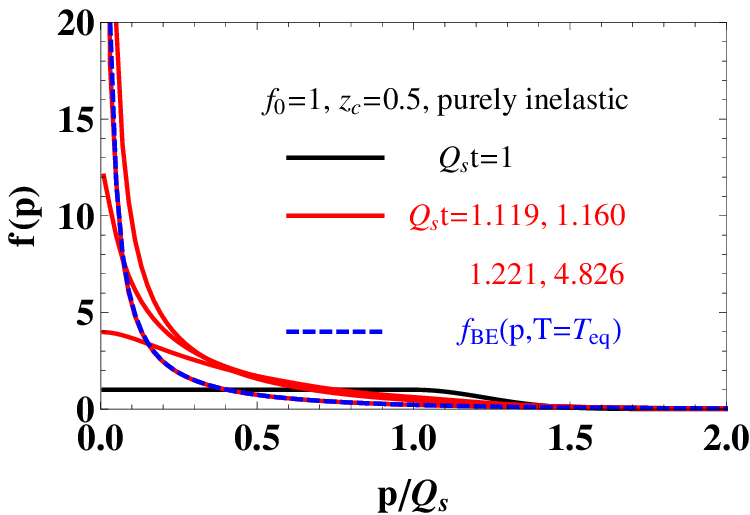}\;\;
\includegraphics[width=5.55cm]{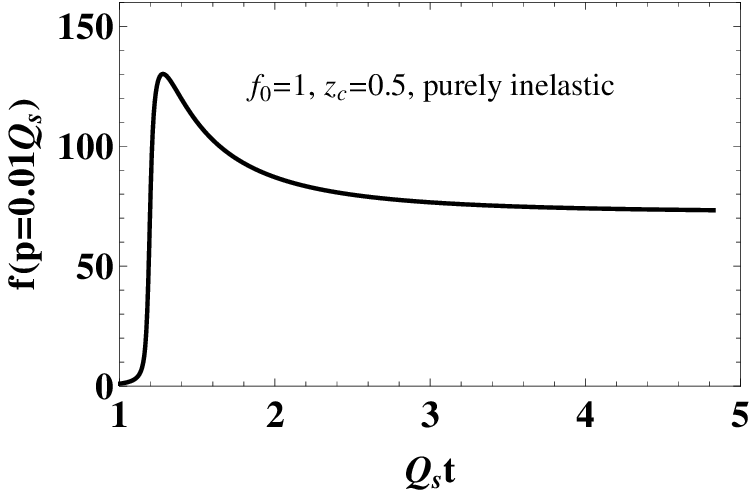}
\caption{(Left) The distribution function $f(p)$ at different time moments during the evolution for purely inelastic collisions;
(Right) The occupation at the smallest grid point $f(p=0.01Q_s)$ as a function of time for purely inelastic collisions.}
\label{fp_inf}
\end{center}
\vspace{-0.7cm}
\end{figure}

In \fig{fp_inf} (left panel), we show the distribution function $f(p)$ at various time moments. It can be  seen that the $f(p)$, starting from the initial Glasma-type shape, smoothly evolves  into a Bose-Einstein distribution after about $Q_st\approx 5$, with a temperature being the supposed value required by energy conservation and a vanishing chemical potential. Very different from the evolution driven by elastic kernel with the same initial condition (see Ref.~\cite{BLM}), in the present purely inelastic case, the system is simply thermalized and during the thermalization there appears no onset of  singularity (Bose-Einstein condensate) in the distribution. This can be explicitly checked by looking at the occupation at the smallest grid point $f(p=0.01Q_s)$ as a function of time (\fig{fp_inf} right panel): its value has a transient behavior of rapid rise and fall and then settles to the supposed thermal value. Clearly, even with overpopulated initial condition, the inelastic process alone does not generate a dynamical onset of Bose-Einstein condensate, as one may reasonably expect.

\begin{figure}
\begin{center}
\includegraphics[width=5.55cm]{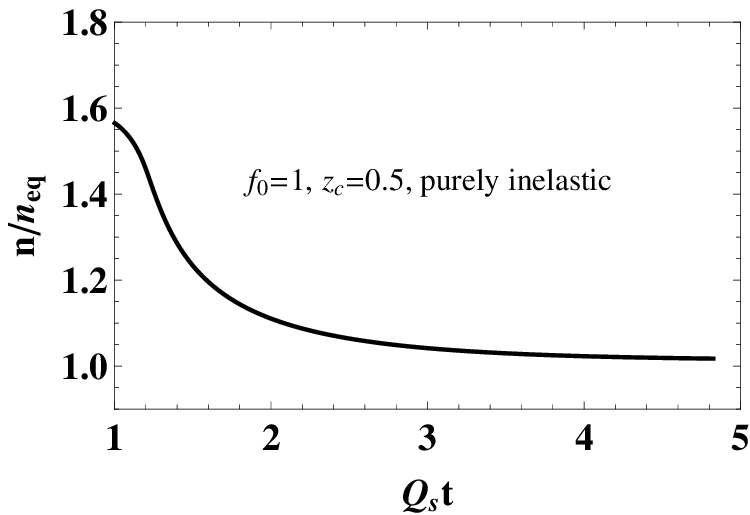}\;\;
\includegraphics[width=5.55cm]{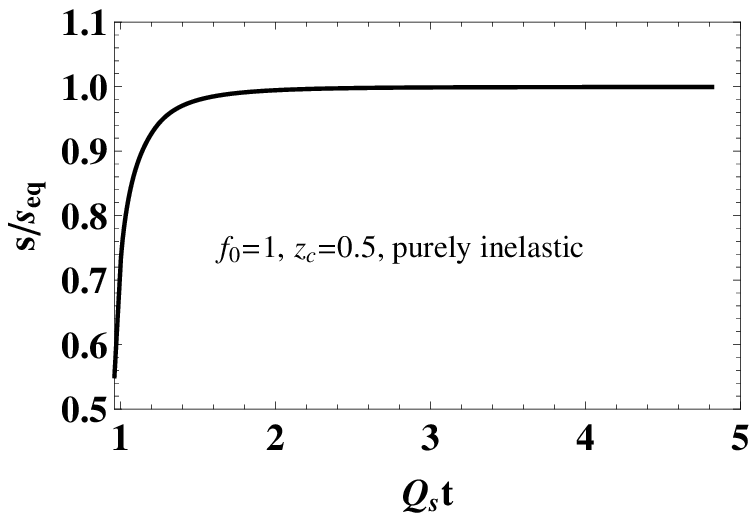}
\caption{The number density $n$ (left) and entropy density $s$ (right), both normalized by the corresponding equilibrium values, as a function of time for purely inelastic collisions.}
\label{den_inf}
\end{center}
\vspace{-0.7cm}
\end{figure}

We have also studied the evolution of global quantities, with the results shown in \fig{den_inf} for the number density (left panel) as well as the entropy density (right panel). The initial high overpopulation in gluon number is efficiently reduced by the inelastic processes, and the number density drops toward the supposed thermal value  determined by equilibrium temperature. The entropy density on the other hand grows rapidly and approaches the thermal value as well. Again all these features provide clear indication that with the purely inelastic kernel the system is simply thermalized as it should be.

From this study, we conclude that the $1\leftrightarrow2$  inelastic processes  as described by our derived kernel (\ref{kernel_12}) thermalize the system efficiently and eliminate excessive gluons from initial conditions effectively, and by these processes {\it alone} no dynamic onset of BEC is to occur. With such benchmark case understood, it is thus tempting to see what will happen when the elastic $2\leftrightarrow2$ processes are also included in addition to the inelastic. As we will show in the next subsections, the evolution of the system gets dramatically changed.

\subsection{From overpopulation toward onset of condensation }

Let us now briefly summarize the kinetic evolution from overpopulation toward onset of condensation in the purely elastic case, as reported in \cite{BLM,Blaizot:2012qd,Liao:2012qk}. The elastic kernel can be re-written in terms of flux as
\begin{eqnarray}
C_{2 \leftrightarrow 2} = - \frac{1}{p^2} \partial_p (p^2\, S(p)) , \,\, S(p) \equiv - \left[ I_a \partial_p f_p + I_b f_p (1+f_p) \right]
\end{eqnarray}
The strong overpopulation leads to a particle flux cascade toward the infrared regime. Analysis of the small $p$ regime shows that it will quickly develop a local thermal form $f^* (p) = 1/[e^{(p-\mu^*)/T^*}-1]$ with $T^*=I_a / I_b$, and the incoming flux will drive the (negative) $\mu^*$ to eventually vanish and reach the onset of a dynamical condensation. This picture is numerically verified in great details in \cite{BLM}. Our discussion of the onset of condensation will stay in this picture (as the elastic term is still present and its flux drives the small $p$ behavior) and we will study how the inelastic process modifies such onset dynamics. It should be emphasize that a vanishing chemical potential $\mu^*$ alone does {\it not} necessarily lead to onset of BEC as is evident from our study of the purely inelastic case in the previous subsection. It is both the vanishing of $\mu^*$ and an elastic-driven divergent flux toward $\vec p=0$ together that would signal the onset of condensation as shown in~\cite{BLM,Scardina:2014gxa,Xu:2014ega}.

\begin{figure}
\begin{center}
\includegraphics[width=5.55cm]{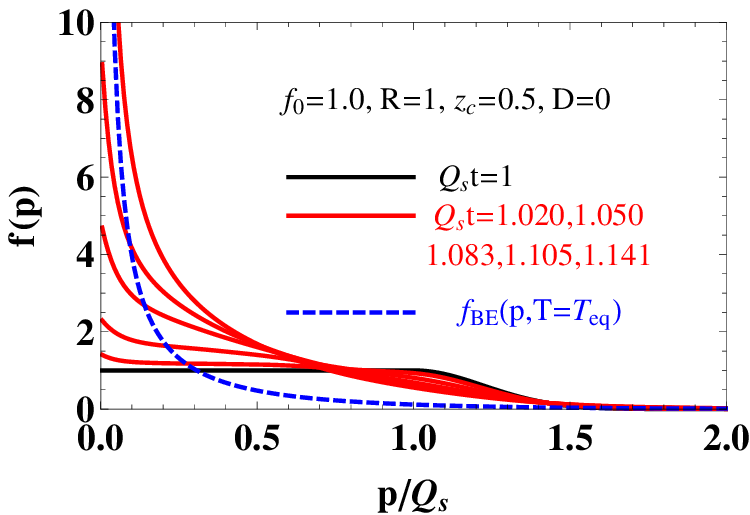}\;\;
\includegraphics[width=5.55cm]{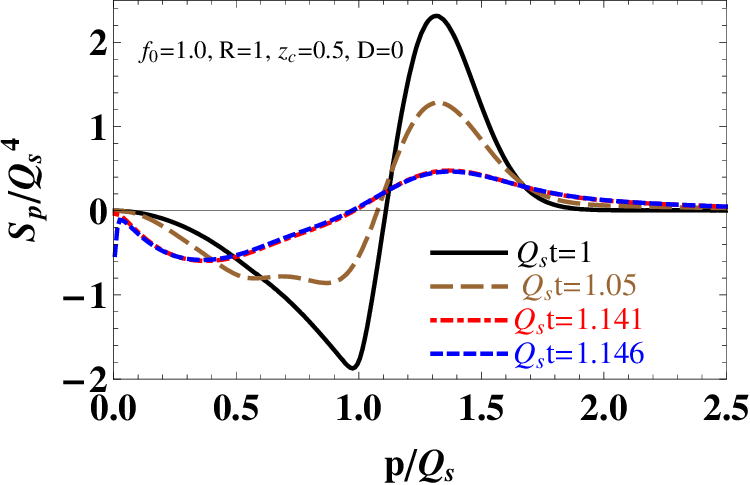}
\caption{(Left) The distribution function $f(p)$ at different time moments during evolution;
(Right) The flux $S(p)$ defined in elastic kernel at different time moments during evolution.}
\label{fz1_fp}
\end{center}
\vspace{-0.7cm}
\end{figure}

Starting with the overpopulated initial condition (\ref{initial}) we have numerically evolved the kinetic equations (\ref{kernel_all}), (\ref{kernel_22}), and (\ref{kernel_12}) with given set of parameters. Shown in \fig{fz1_fp} is the solution with $R=1$. In \fig{fz1_fp} left we show the distribution function $f_p$ at different time moments, and one can see that even with the presence of inelastic term, the small $p$ part of the distribution is quickly filled up and becomes a local thermal form $f^* (p) = 1/[e^{(p-\mu^*)/T^*}-1]$ despite that the distribution in the wide range of (bigger) momentum region is still far from equilibrium shape, and the small $p$ part becomes steeper and steeper with time (meaning decreasing $|\mu^*|$).
In \fig{fz1_fp} right we show the corresponding flux $S(p)$ from the elastic kernel. Again the flux behaves very similarly to the purely elastic case: one see a linear behavior at small p, $S\propto - p$ and eventually upon onset of condensation the flux diverges (see the blue curve near $p=0$).

\begin{figure}
\begin{center}
\includegraphics[width=5.55cm]{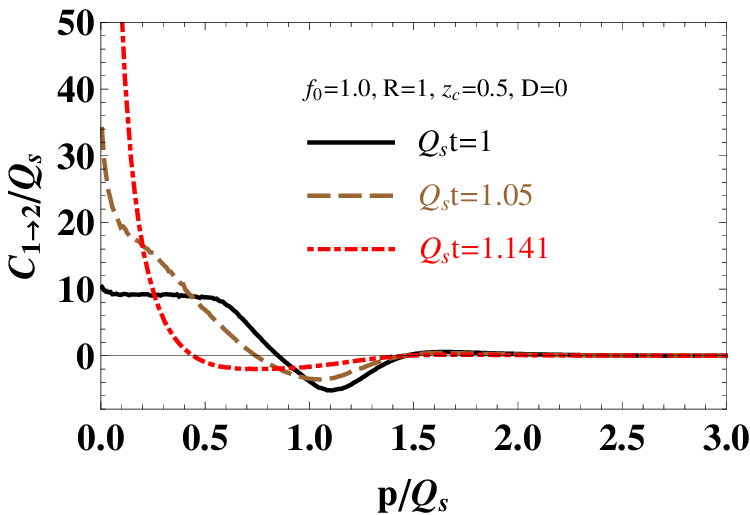}\;\;
\includegraphics[width=6.05cm]{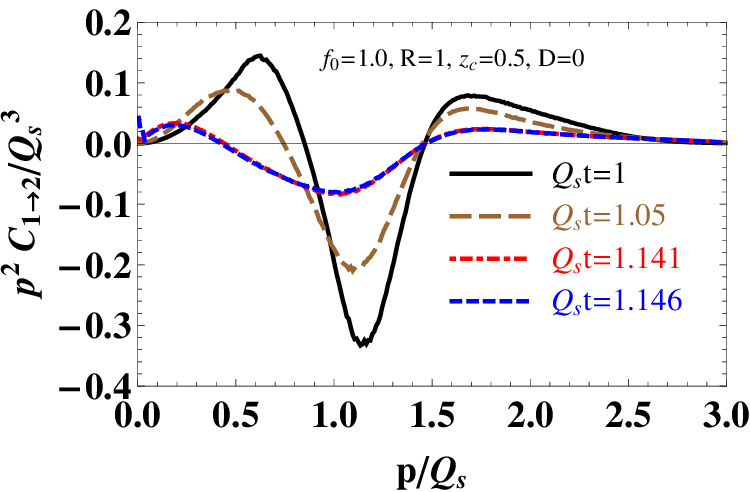}
\caption{The inelastic kernel $C_{1 \leftrightarrow 2}(p)$ (Left) and $p^2 C_{1 \leftrightarrow 2}(p)$ (Right) at different time moments during evolution.}
\label{fz1_C12}
\end{center}
\vspace{-0.7cm}
\end{figure}

To get an intuitive idea of the contribution of the inelastic kernel, we plot the $C_{1 \leftrightarrow 2}$ and $p^2 C_{1 \leftrightarrow 2}$ in \fig{fz1_C12}. One can see that the kernel is large and positive at small $p$, small and positive at large $p$, while negative at intermediate $p$. This could be qualitatively understood: significant number of particles with intermediate momenta will merge toward high momenta and split toward low momenta that will fill up UV and IR region while decrease the occupation at intermediate momenta. We also notice that upon onset (the blue dashed curve in the right panel) the inelastic kernel near $p=0$ shows a divergent behavior in consistency with the elastic flux behavior.

\begin{figure}
\begin{center}
\includegraphics[width=5.55cm]{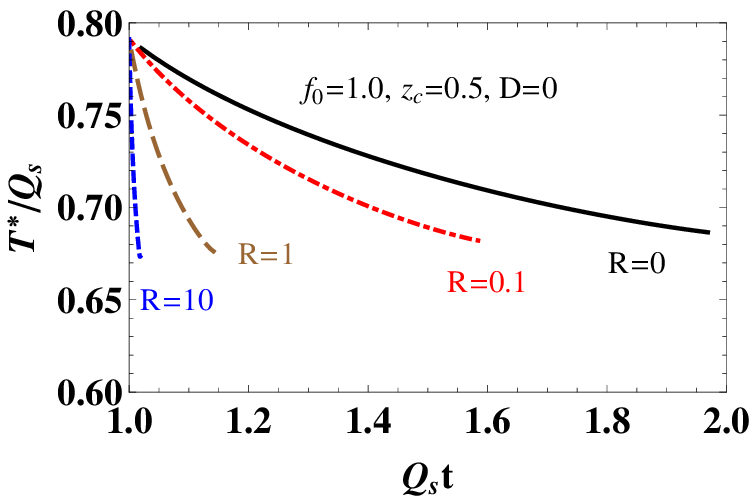}\;\;
\includegraphics[width=5.55cm]{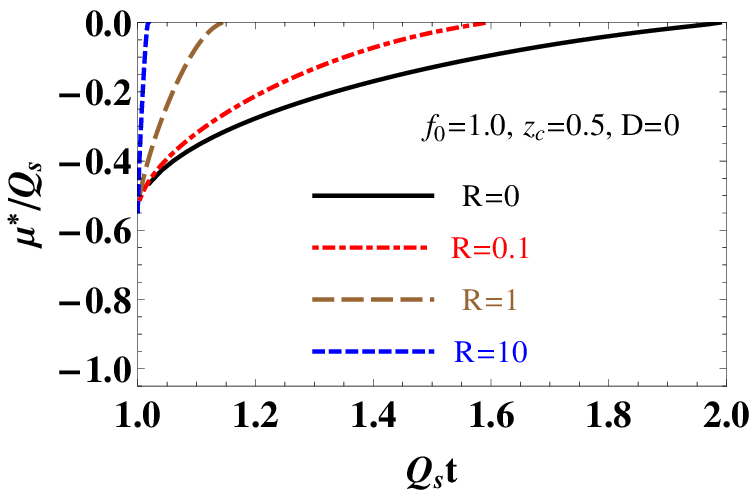}
\caption{ The local thermal form parameters $T^*$ (Left) and $\mu^*$ (Right) as functions of time for different values of parameter $R$.}
\label{fz1_TMu}
\end{center}
\vspace{-0.7cm}
\end{figure}

One can directly examine the locally determined $T^*$ and $\mu^*$ (see \cite{BLM} for details) at each time moment during the evolution: these results are shown in \fig{fz1_TMu}. Here we also compare the results for different strength of the inelastic collision $R=0,0.1,1,10$ (noting that the $R=0$ case corresponds to purely elastic collision). In all cases we can see that the local ``chemical potential'' $\mu^*$ decreases rather rapidly toward zero. We also show the distribution $f(p)$ at the smallest grid point in our calculation $p=0.005Q_s$ as a function of time in \fig{fz1_thermal}, which shows very rapid increase of the occupation in consistence with the vanishing of $\mu^*$. What is most striking is that with increase values of $R$ this evolution toward the onset of condensation $\mu^* \to 0$ becomes faster and faster. The $R=1$ case is already much faster than the purely elastic case. This is to say, contrary to expectation that the inelastic process may ``kill'' the
 strong overpopulation quickly,  the existence of inelastic collision actually  {\bf speeds up}  significantly the process of populating the infrared regime and building up a local thermal form with vanishing $\mu^*$, which when combined with the structure of elastic kernel will then lead to the onset of condensation.

\subsection{Small $p$ analysis of the inelastic kernel}

\begin{figure}
\begin{center}
\includegraphics[width=5.55cm]{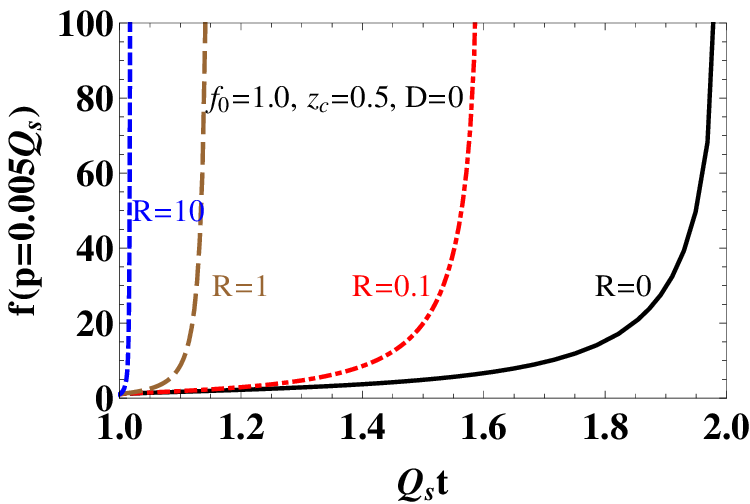}\;\;
\caption{The occupation at the smallest grid point $f(p=0.005Q_s)$ as a function of time for different values of parameter $R$.}
\label{fz1_thermal}
\end{center}
\vspace{0.cm}
\end{figure}

To understand better the influence of inelastic collision on the small $p$ region, let us examine the kernel (\ref{kernel_12}) for $p\to 0$ limit before the onset of condensation. Provided that $f_0=f(p=0)<\infty$ and that its derivatives with respect to $p$ at $p=0$ is also finite, we can have the expansion $f(p\to 0)\approx f_0 + f'_0\, p + ...$. If we look at a small enough $p$ regime, then all the involved momenta ($p, zp, (1-z)p, p/(1-z),pz/(1-z)$) in the kernel  (\ref{kernel_12}) can be considered small and we can use the expansion for them. This leads to
\begin{eqnarray}
{\cal C}^{\rm eff}_{1\leftrightarrow 2} (p\to 0) \to
R\frac{I_a}{I_b}\ls A_0 f_0(1+f_0)+ A_1 f'_0(1+2f_0) \, p+\hat{O}(p^2)\rs,
\end{eqnarray}
where we have introduced the constants
\begin{eqnarray}
A_0&=&\ln\frac{1}{1-z_c}+\frac{1}{6}\frac{z_c(11z_c^2-27z_c+18)}{(1-z_c)^3},\non
A_1&=&\ln\frac{1}{1-z_c}-\frac{1}{12}\frac{z_c(25z_c^3-88z_c^2+108z_c-48)}{(1-z_c)^4}.
\end{eqnarray}
All these $A$'s are positive for $0<z_c<1$. Clearly for  sufficiently small $p$ the leading term in the inelastic kernel $\sim f_0(1+f_0) A_0$ is {\it always positive} and becomes bigger and bigger with increasing $f_0$ (which is a kind of ``self-amplification''). This will tend to increase the particle number near $p=0$ very rapidly and the effect becomes stronger with increasing values of $R$, which explains the behavior seen in \fig{fz1_thermal}.

Physically this behavior may be understood in two ways. First note that the inelastic kernel has its fixed point to be $1/(e^{p/T}-1)$ which at small $p$ is $\sim 1/p $ so as long as $f(p=0)$ is finite yet the inelastic kernel will try to fill it up toward $1/p$.  Second, this is also related to the Boson nature: if all involved particles are from small $p$, then the merging rate is like $\sim f_0^2 (1+f_0)$ while the splitting rate is like $\sim f_0 (1+f_0)^2$ so the splitting ``wins'' due to Bose enhancement for the final state particles and it increases particle number at small $p$.
To conclude, the inelastic kernel contribution is always positive at very small $p$ and it will catalyze and speed up the onset of a Bose condensation (which itself is driven by the elastic term at $\mu^* \to 0$).

\begin{figure}
\begin{center}
\includegraphics[width=5.55cm]{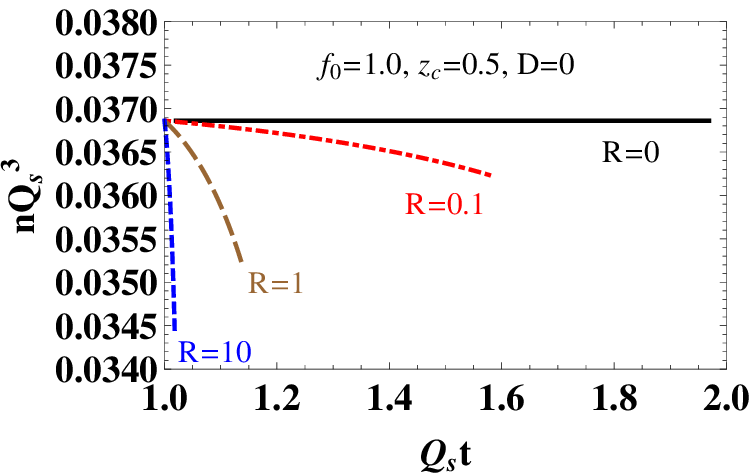}\;\;
\includegraphics[width=5.55cm]{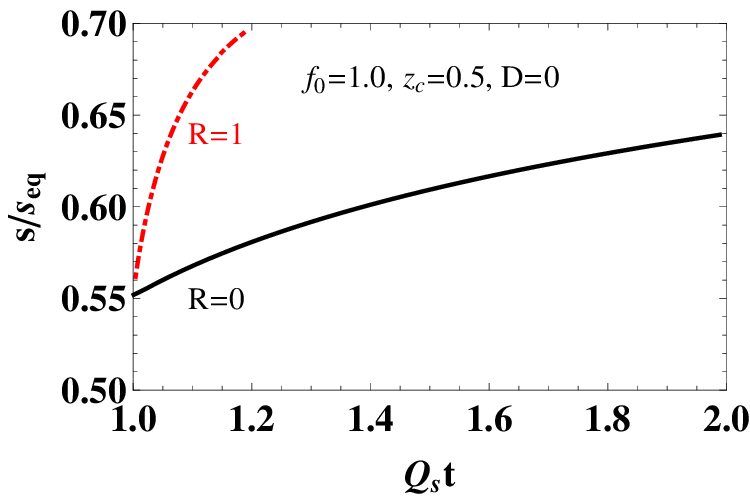}
\caption{(Left) The particle number density as a function of time for different values of parameter $R$.
(Right)  The entropy  density as a function of time for  $R=1$ and $R=0$.}
\label{fz1_density}
\end{center}
\vspace{-0.7cm}
\end{figure}

\subsection{Change of particle number from inelastic kernel}

While the inelastic kernel always increases the occupation at sufficiently small $p$, it may still decrease the the total particle number. Indeed as shown in \fig{fz1_density} (left panel), the total particle number decreases when $R>0$, and it decreases more rapidly for larger $R$.

To understand the change of particle number $n=\int d^3\bp/(2\p)^3 f(p)$, one can integrate the two sides of the kinetic equation (\ref{kernel_all}) and obtain
\begin{eqnarray}
&& \pt_t n(t)
= R\frac{I_a}{2\p^2I_b}
\int dp p^2\int_{0}^{z_c} \frac{dz}{1-z}
\ls f_pg_{(1-z)p}g_{zp}-g_pf_{(1-z)p}f_{zp}\rs \nonumber \\
&& = R\frac{I_a}{2\p^2I_b}
\int dp p^2\int_{0}^{z_c} \frac{dz}{1-z}
\ls    f_p + f_p f_{(1-z)p} +f_p f_{zp} - f_{(1-z)p}f_{zp}\rs.
\end{eqnarray}
From the above one can see the for the region $z\to 0$  the leading order in the $z$-integrand becomes $\sim f_p (1+f_p)$ and the contribution is positive, i.e. increasing particle number. For general $z$, the $z$-integrand can be rewritten as
\begin{eqnarray}
 \pt_t n(t)  = R\frac{I_a}{2\p^2I_b}
\int dp p^2\int_{0}^{z_c} \frac{dz}{1-z}
\ls    f_p (1+f_p)  - ( f_{(1-z)p} -f_p) ( f_{zp} - f_p) \rs.
\end{eqnarray}
We see that for not too small $z$, the momenta $zp$, $(1-z)p$ become well separated from $p$ and the second term in the above integrand becomes important and its contribution is negative which decreases the particle number.

In \fig{fz1_density} (right panel) we also show the entropy density as a function of time and compare the case with $R=1$ and the purely elastic case with $R=0$. One can see that with inelastic collision included, the entropy density increases much faster. That is, the inelastic process tends to accelerate the thermalization.

\begin{figure}
\begin{center}
\includegraphics[width=5.55cm]{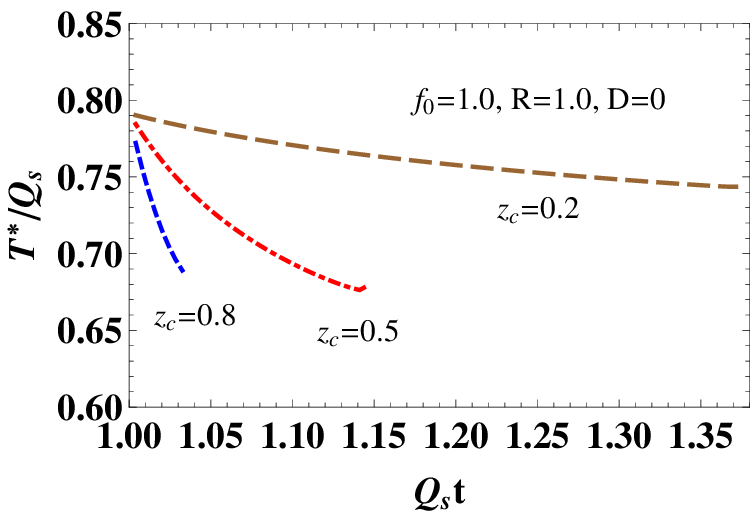}\;\;
\includegraphics[width=5.55cm]{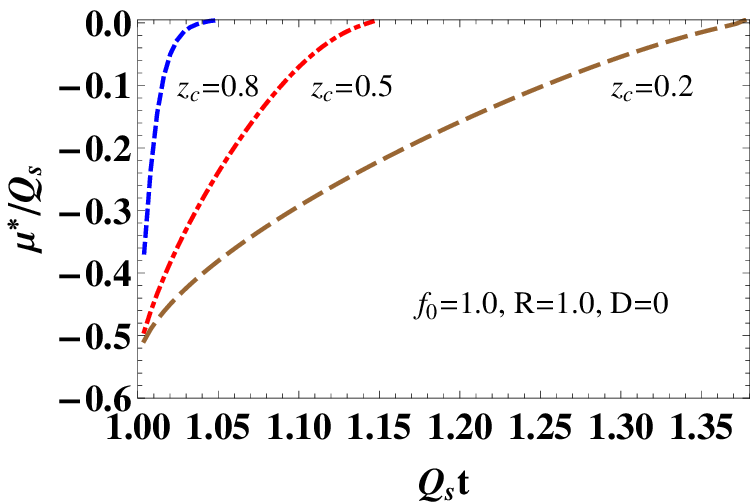}
\caption{The local thermal form parameters $T^*$ (Left) and $\mu^*$ (Right) as functions of time for $z_c=0.2,0.5,0.8$ respectively.}
\label{fz1_varyz1}
\end{center}
\vspace{-0.7cm}
\end{figure}

\subsection{Dependence on the parameter $z_c$}

\begin{figure}
\begin{center}
\includegraphics[width=5.55cm]{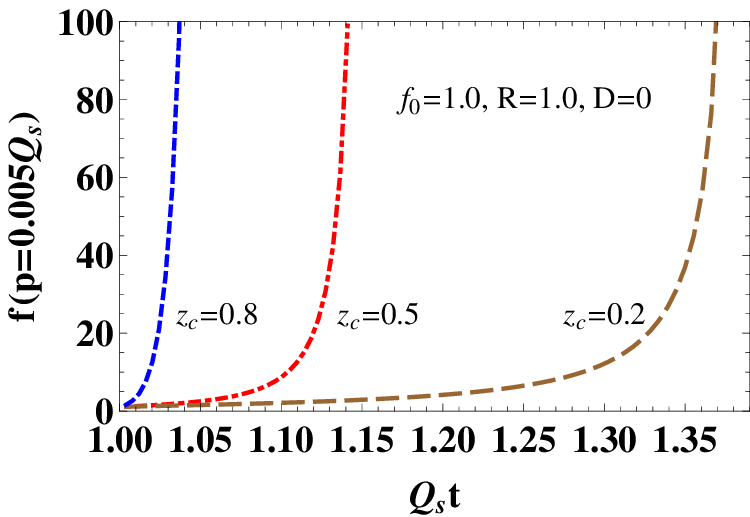}\;\;
\includegraphics[width=5.55cm]{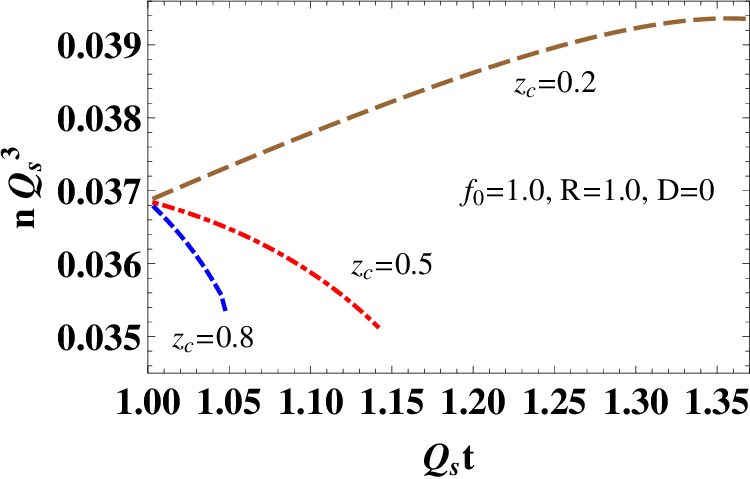}
\caption{The occupation at the smallest grid point $f(p=0.005Q_s)$ (Left) and the particle number density (Right) as functions of time for  $z_c=0.2,0.5,0.8$ respectively}
\label{fz1_varyz2}
\end{center}
\vspace{-0.7cm}
\end{figure}

Finally we study the dependence on the parameter $z_c$ which is the kinematic cut to make sure the validity of the approximations used for the matrix element. Generally speaking, with larger $z_c$ we include more effects from the inelastic process. To see how the results depend on $z_c$, we fix other parameters and compare the results for different choices of $z_c$. In \fig{fz1_varyz1}, we show the local thermal form parameters $T^*$ (left panel) and $\mu^*$ (right panel) as functions of time for $z_c=0.2,0.5,0.8$ respectively. In \fig{fz1_varyz2}, we show the occupation at the smallest grid point (left panel) and the total particle number (right panel) as functions of time for  $z_c=0.2,0.5,0.8$ respectively. From the plots we can see that indeed with larger $z_c$ the $f_0$ increases faster and $\mu^*$ vanishes faster as expected for stronger inelastic effect. For the particle number, the case with $z_c =0.2$ actually has $n$ increasing with time, which can be understood from the analysis in the previous subsection. The particle number in both $z_c=0.5$ and $z_c=0.8$ cases drops with time and does so faster for larger $z_c$.  In passing let us mention that we have also studied the dependence on the parameter $D$: basically increasing $D$ will enhance the effect from elastic collision and also speed up the thermalization in general, as well as reach onset of condensation at earlier times compared with $D=0$ case.

\section{Conclusion}

In summary, we have studied the kinetic evolution of a highly overpopulated system starting from Glasma-type initial condition with the presence of both elastic and inelastic collisions. Using the Gunion-Bertsch formula for the $2\leftrightarrow 3$ matrix element, we have derived the inelastic collision kernel under the collinear and small angle approximations. Putting together the inelastic kernel together with the previously obtained elastic kernel, we have then numerically solved the kinetic evolution for varied choices of parameters. Our main finding is that the inelastic process has two effects: globally changing (mostly reducing) the total particle number, while locally at small $p$ always filling up the infrared regime extremely quickly. The latter effect is shown both from numerics and by analytic analysis. This effect significantly speeds up the emergence of local thermal form near $p=0$ and the vanishing of local ``chemical potential'' $\mu^*$ as previously found in
  the purely elastic collision case to lead to the onset of dynamical Bose condensation. Therefore in our present approach of including the inelastic scattering, we conclude that, contrary to some previously discussed expectations about the role of number non-conserving processes, the inelastic collision actually helps to build up the local ``critical form'' $\sim 1/p$ much faster and catalyzes the onset of condensation in the overpopulated Glasma.
\begin{figure}
\begin{center}
\includegraphics[width=6.6cm]{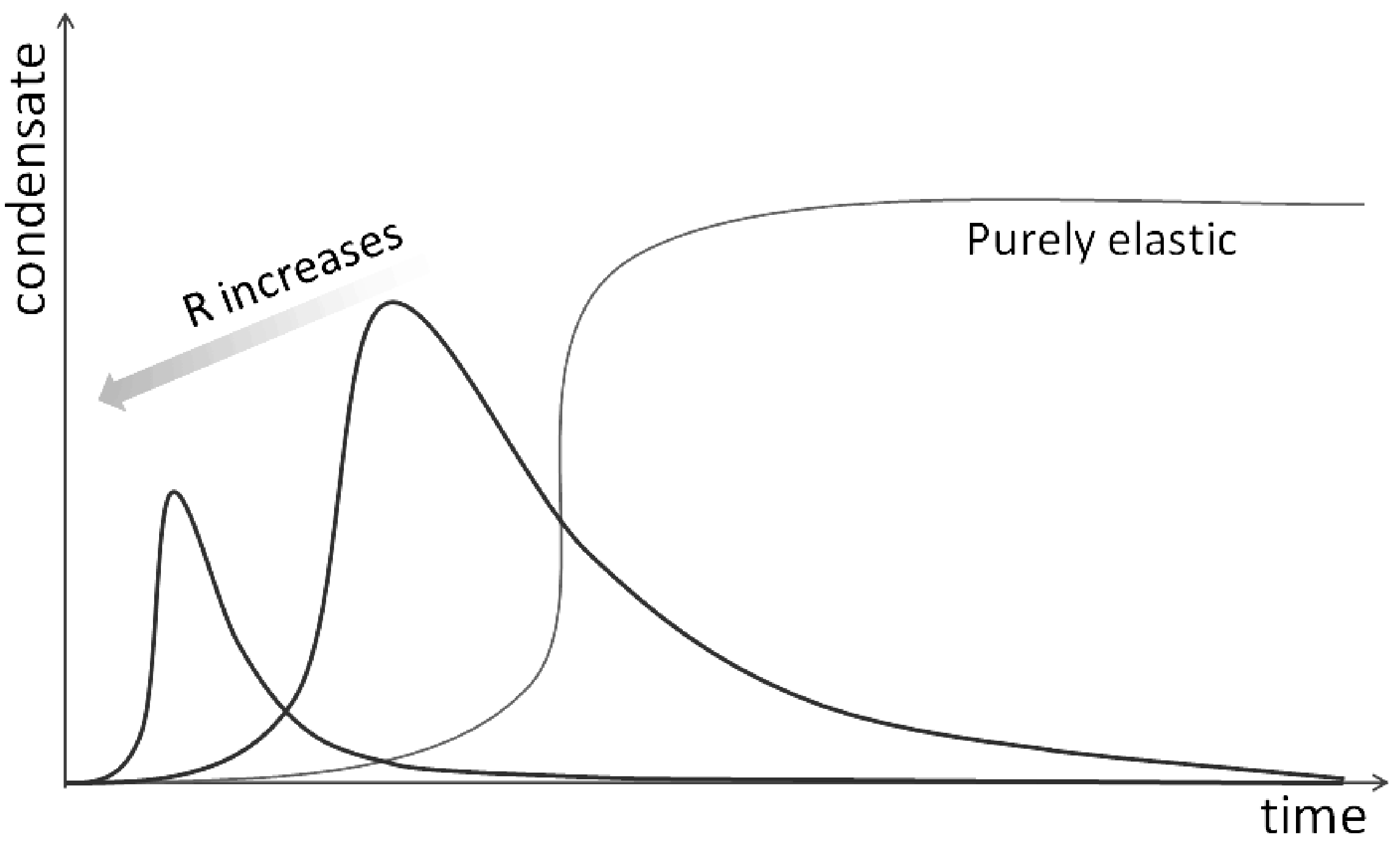}
\caption{A conjectured evolution of the condensate.}
\label{conje}
\end{center}
\vspace{-.6cm}
\end{figure}

Our finding may sound counter-intuitive at first, as the usual conception would suggest that increasing the strength of the inelastic collisions tends to obstruct more effectively the formation of any condensate. It should however be emphasized that the evolution toward onset of BEC that has been studied thus far is not the end of the story. Our analysis addresses the evolution up to the onset of BEC while does not treat the evolution afterwards. As is well known in the BEC literature (see e.g. \cite{Semikoz:1994zp,Semikoz:1995rd}), in order to describe the kinetic evolution of the system with the presence of condensate, a new set of kinetic equations is needed  for an explicit description of the coupled evolution for a condensate plus a regular distribution. Efforts are underway to derive these equations, and so far a kinetic study of the stage after BEC onset for the Glasma system has not been achieved to our best knowledge. However, it appears very plausible that the subsequent evolutions may develop as follows: immediately after onset, the strong IR flux will not cease right away but continue for a while and thus drive the condensate to grow in time; at certain point, the time would be long enough to allow the inelastic processes to decrease the total number density adequately and cause the condensate to decay thus decreasing in time; eventually the inelastic processes will be able to remove all excess gluons and lead to the thermal equilibrium state with neither condensate nor any chemical potential. While the detailed understanding of such dynamic processes can only be achieved  through solving the new set of kinetic equations, one can reasonably expect that with increasing strength of the inelastic processes the whole evolution would be faster. Thus the following overall picture may  likely be the case: with increasing strength, the inelastic processes on one hand catalyze the onset of condensation initially, while on the other hand  eliminate the fully formed condensate faster, thus limiting the time duration for the presence of  condensate to be shorter. A schematic picture of such conjectured full evolution is shown in \fig{conje}, which is in line with the usual conception. It is worth mentioning that recent analysis in \cite{Zhang:2012vi} has shown that the the $2\leftrightarrow3$ inelastic cross section from exact matrix element becomes significantly smaller than that from the Gunion-Bertsch formula, and amounts to $\sim 20\%$ of the $2\leftrightarrow2$ cross section. It therefore seems very plausible that a realistic choice of $R$ value would be rather modest, which may imply a considerable time window for the condensate to be sizable and play an important role for the evolution. A complete investigation of the evolution including the  condensate will be future project to be reported elsewhere. Furthermore how medium effects like the screening as well as the Landau-Pomeranchuk-Migdal effect may influence the glasma evolution deserves a careful study in the kinetic framework as well~\cite{Arnold:2002ja,Arnold:2002zm} and it will also be a future task.

Lastly, we'd like to mention a recent kinetic theory studies \cite{York:2014wja,Kurkela:2014} that also includes both an elastic kernel and an effective inelastic kernel. The analysis of \cite{York:2014wja,Kurkela:2014} appears to bear different conclusions than ours, regarding the evolution in the very infrared region. Particularly, in contrast to our findings, Refs. \cite{York:2014wja,Kurkela:2014} did not observe the formation of a condensate. It is important to understand the origin of such difference between our study and theirs. A major factor may likely contribute to the different results: while we use the vacuum matrix elements for both elastic and inelastic processes, the authors of \cite{York:2014wja,Kurkela:2014} use medium-modified effective matrix elements for both processes. A comparative study  will be crucial and it is highly desired to address, in both approaches, the following questions: does an overpopulated initial condition with pure elastic kernel lead to BEC onset? does an overpopulated initial condition with pure inelastic kernel thermalize without condensation? whether an overpopulated initial condition will lead to BEC onset or not, when both kernels are included? These will be investigated and reported in a future work.

\section*{Acknowledgements}
The authors are particularly grateful to L. McLerran for very helpful discussions. J.L. also thanks J.-P. Blaizot, F. Gelis, and R. Venugopalan for collaborations and communications that motivated and benefited this work. J.L. is supported by the National Science Foundation (Grant No. PHY-1352368), he also acknowledges the RIKEN BNL Research Center for partial support. X.G.H. is supported by Shanghai Natural Science Foundation (Grant No. 14ZR1403000).

\section*{Appendix}

\numberwithin{equation}{section}
\appendix
\section {$gg\leftrightarrow ggg$ matrix element}\label{amplitude}
\begin{figure}
\begin{center}
\includegraphics[width=5.5cm]{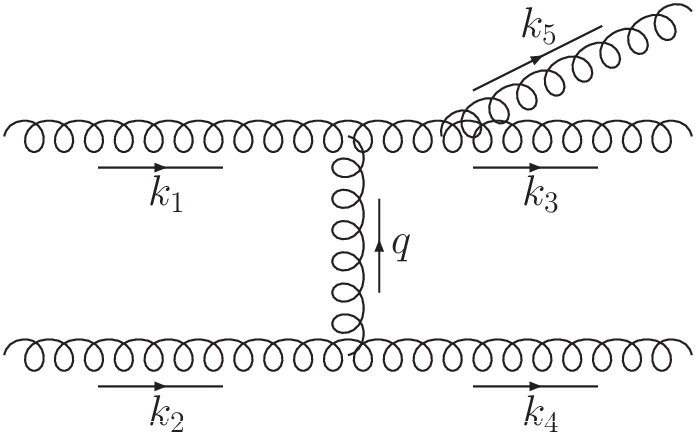}
\caption{A typical Feynman diagram for $M_{2 \leftrightarrow 3}$.}
\label{f1}
\end{center}
\vspace{-0.7cm}
\end{figure}

The invariant $gg\leftrightarrow ggg$ or $2\leftrightarrow3$ (squared) matrix element summed over all final states and also summed over all initial
states is computed by considering 25 different Feynman diagrams~\cite{Berends:1981rb}. We quote it here:
\begin{eqnarray}
\label{M23general}
|M_{2\leftrightarrow3}|^2&=&g^6N_c^3N_g\frac{\cal N}{\cal D}[(12345)+(12354)+(12435)+(12453)+(12534)\non&&+(12543)+
(13245)+(13254)+(13524)+(14235)+(14325)],\non
\end{eqnarray}
where $N_c=3$ is the number of color, $N_g=2(N_c^2-1)$ is the gluon degeneracy number, and other notations are defined as
\begin{eqnarray}
{\cal N}&=&(12)^4+(13)^4+(14)^4+(15)^4+(23)^4+(24)^4\non&&+(25)^4+(34)^4+(35)^4+(45)^4,\non
{\cal D}&=&(12)(13)(14)(15)(23)(24)(25)(34)(35)(45),\non
(ijklm)&=&(ij)(jk)(kl)(lm)(mi),\non
(ij)&\equiv& k_i\cdot k_j.
\end{eqnarray}
Because $|M_{2\leftrightarrow3}|^2$ is completely symmetry in $k_i, i=1-5$, let's take $k_1$ and $k_2$ as
the hard momenta in the entrance channel, $k_3$ and $k_4$ as the hard momenta in
the exit channel, and $k_5$ as the emitted soft gluon. We denote the exchanging momentum
as $q=k_2-k_4$. A typical Feynman diagram illuminating
this setup is shown in \fig{f1}.

Define the Mandelstam variables as~\cite{Abir:2010kc,Bhattacharyya:2011vy}
\begin{eqnarray}
s=(k_1+k_2)^2=2(12),\; t=(k_1-k_3)^2=-2(13),\; u=(k_1-k_4)^2=-2(14),\non
s'=(k_3+k_4)^2=2(34),\; t'=(k_2-k_4)^2=-2(24),\; u'=(k_2-k_3)^2=-2(23).\nonumber
\end{eqnarray}
In addition, the following relations involving $k_5$ hold:
\begin{eqnarray}
(15)=\frac{s+t+u}{2},\; (25)=\frac{s+t'+u'}{2},\; (35)=\frac{s+t'+u}{2},\;(45)=\frac{s+t+u'}{2}.\nonumber
\end{eqnarray}
In terms of the Mandelstam variables, $|M_{2\leftrightarrow3}|^2$ can be written as
\begin{eqnarray}
|M_{2\leftrightarrow3}|^2&=&32g^6N_c^3N_g{\cal N}\Big[\frac{1}{s'(s+u+t)(s+u'+t')}\Big(\frac{1}{tt'}+\frac{1}{uu'}\Big)
\non&&+
\frac{1}{s(s+u'+t)(s+u+t')}\Big(\frac{1}{tt'}+\frac{1}{uu'}\Big)\non&&-
\frac{1}{t'(s+u+t)(s+u+t')}\Big(\frac{1}{ss'}+\frac{1}{uu'}\Big)\non&&-
\frac{1}{t(s+u'+t)(s+u'+t')}\Big(\frac{1}{ss'}+\frac{1}{uu'}\Big)\non&&-
\frac{1}{u'(s+u+t)(s+u'+t)}\Big(\frac{1}{tt'}+\frac{1}{ss'}\Big)\non&&-
\frac{1}{u(s+u+t')(s+u'+t')}\Big(\frac{1}{tt'}+\frac{1}{ss'}\Big)\Big],
\end{eqnarray}
where
\begin{eqnarray}
{\cal N}&=&\frac{1}{16}[s^4+t^4+u^4+s'^4+t'^4+u'^4+(s+u+t)^4\non&&+(s+u'+t')^4+(s+t'+u)^4+(s+t+u')^4].\nonumber
\end{eqnarray}
Because $|M_{2\leftrightarrow3}|^2$ is very singular when, for example, $t,t'\ra 0$, we can expand it around these singularities order by order in
some small momenta.
To this end, let's assume that the exchanging momentum $q=k_2-k_4$ is small ($k_5$ is also small).
In this case $t,t'$ are small, while $s,s',u,u'$ are large,
and $-u\ra -u'\ra s'\ra s$. With other choices of picking the small exchanging momenta and soft emitted momenta, we can get other while
equivalent expansions. We will come to this point latter. Keeping only leading order and subleading order
terms in $q$ and $k_5$, we have
\begin{eqnarray}
s'&=&(k_1+k_2-k_5)^2=s-2(k_1+k_2)\cdot k_5+O(k_5^2),\non
u&=&-s+2k_1\cdot k_5+O(q^2),\non
u'&=&-s+2k_2\cdot k_5+O(q-k_5)^2,\non
t&=&(q-k_5)^2,\non
t'&=&q^2.\nonumber
\end{eqnarray}
In addition, we have
\begin{eqnarray}
k_2\cdot(k_4+q)=0\Rightarrow k_2\cdot q=-k_2\cdot k_4=\frac{1}{2}(k_2-k_4)^2=\frac{1}{2}q^2,\non
t-t'=-2q\cdot k_5\Rightarrow-2q\cdot k_2-2k_1\cdot(q-k_5)=2k_2\cdot k_4-2k_1\cdot k_3=t-t'=-2q\cdot k_5
\non\Rightarrow k_1\cdot(k_5-q)=\frac{q^2}{2}-q\cdot k_5.\nonumber
\end{eqnarray}
Then ${\cal N}$ and $|M_{2\leftrightarrow3}|^2$ can be simplified as
\begin{eqnarray}
{\cal N}&=&\frac{s^4}{4}-s^3(k_2\cdot k_5+k_1\cdot k_5)+O(k_5^2/s,q^2/s,k_5\cdot q/s),\\
\label{M23}
|M_{2\leftrightarrow3}|^2&=&32g^6N_c^3N_g\frac{{\cal N}}{tt'}\Big[\frac{1}{s'(s+u+t)(s+u'+t')}+
\frac{1}{s(s+u'+t)(s+u+t')}\non&&-
\frac{1}{u'(s+u+t)(s+u'+t)}-
\frac{1}{u(s+u+t')(s+u'+t')}\Big]+O\lb\frac{t}{s}\rb\non
&=&32g^6N_c^3N_g\frac{{\cal N}}{tt'}\frac{s+(k_1+k_2)\cdot k_5}{s^2(k_1\cdot k_5)(k_2\cdot k_5)}+O\lb \frac{k_5^2}{s},\frac{q^2}{s},\frac{k_5\cdot q}{s}\rb\non
&=&8g^6N_c^3N_g\frac{s^2}{tt'}\frac{s-3(k_1+k_2)\cdot k_5}{(k_1\cdot k_5)(k_2\cdot k_5)}+O\lb \frac{k_5^2}{s},\frac{q^2}{s},\frac{k_5\cdot q}{s}\rb\non
&=&32g^6N_c^3N_g\frac{(k_1\cdot k_2)^2}{q^2(q-k_5)^2}\frac{2k_1\cdot k_2-3(k_1+k_2)\cdot k_5}{(k_1\cdot k_5)(k_2\cdot k_5)}+O\lb \frac{k_5^2}{s},\frac{q^2}{s},\frac{k_5\cdot q}{s}\rb.\non
\end{eqnarray}
In the center-of-mass frame of $k_1$ and $k_2$, it goes to
\begin{eqnarray}
|M^{c.m.}_{2\leftrightarrow3}|^2&\approx&|M_{\rm GB}|^2\lb1-3\frac{|\bk_5|}{\sqrt{s}}\rb
\lb1+\frac{q_0^2-\bq_\parallel^2}{\bq_\perp^2}\rb
\lb1+\frac{(q_0-k_{50})^2-(\bq_\parallel-\bk_{5\parallel})^2}{(\bq_\perp-\bk_{5\perp})^2}\rb\non
&\approx&|M_{\rm GB}|^2\lb1-3\frac{|\bk_{5\perp}|}{\sqrt{s}}\rb,
\end{eqnarray}
where $\bq_\parallel=(\bq\cdot\bv_1)\bv_1$ and
\begin{eqnarray}
\label{MGB}
|M_{\rm GB}|^2=
32g^6N_c^3N_g\frac{s^2}{\bq_\perp^2(\bq_\perp-\bk_{5\perp})^2\bk_{5\perp}^2}
\end{eqnarray}
is the Gunion-Bertsch formula~\cite{Gunion:1981qs}. Here we have used the fact that $q_0^2=(\bv_2\cdot\bq)^2+O(\bq^2_\perp q_0/|\bk_2|)$ and $(q_0-k_{50})^2=[\bv_1\cdot(\bq-\bk_5)]^2+O[(\bq_\perp-\bk_{5\perp})^2(q_0-k_{50})/|\bk_1|]$ for soft $q$ and $k_5$.
Thus, the Gunion-Bertsch formula is the leading order result in soft $q$ and $k$ expansion; and \eq{M23} is the result including both the leading (Gunion-Bertsch) and the
subleading order terms. Higher order terms can also be obtained, but we will not use them.
Note that one can naively boosts the Gunion-Bertsch formula from the center-of-mass frame to
a general frame by using the replacements $\bk_{5\perp}^2\ra4(k_1\cdot k_5)(k_2\cdot k_5)/s$,
$\bq_\perp^2\ra4(k_1\cdot k_4)(k_2\cdot k_4)/s$, and $(\bq_\perp-\bk_{5\perp})^2\ra4(k_1\cdot k_3)(k_2\cdot k_3)/s$:
\begin{eqnarray}
\label{MGBgeneral}
|M_{\rm GB}|^2=
16g^6N_c^3N_g\frac{(k_1\cdot k_2)^5}{(k_1\cdot k_3)(k_2\cdot k_3)(k_1\cdot k_4)(k_2\cdot k_4)(k_1\cdot k_5)(k_2\cdot k_5)}.
\end{eqnarray}
This expression coincides with \eq{M23} at leading order but not at next to leading order in soft $q$ and $k_5$ expansion.

\section {The collision kernel ${\cal C}_{2\leftrightarrow3}$ at collinear approximation}
\label{colli}
The collision kernel ${\cal C}_{2\leftrightarrow3}$ has a very complicated structure,
in this section, we simplify it by taking the collinear approximation,
i.e., $\bv_k\simeq\bv_1$ or $\bv_k\simeq\bv_p$.

We rewrite the collision kernels ${\cal C}^a_{2\leftrightarrow3}$ and ${\cal C}^b_{2\leftrightarrow3}$ as (We denote $1+f_i$ by $g_i$)
\begin{eqnarray}
{\cal C}^a_{2\leftrightarrow3}&=&\frac{1}{N_g}\int_{123kl}\frac{2E_l}{2E_p}\int\frac{d^4q}{(2\p)^4}\int\frac{dl_0}
{2\p}(2\p)^4\d^4(p+q-l)(2\p)^4\d^4(p_1-q-p_2)\non&&\times(2\p)^4
\d^4(l-k-p_3)|M_{1p\leftrightarrow23k}^a|^2
(g_pg_1f_2f_3f_k-f_pf_1g_2g_3g_k),\non
{\cal C}^b_{2\leftrightarrow3}&=&\frac{1}{N_g}\int_{123kl}\frac{2E_l}{2E_p}\int\frac{d^4q}{(2\p)^4}\int\frac{dl_0}
{2\p}(2\p)^4\d^4(p+q-l)(2\p)^4\d^4(p_1-q-p_2)\non&&\times(2\p)^4
\d^4(l+k-p_3)|M_{23\leftrightarrow1pk}^b|^2
(g_pg_1g_kf_2f_3-f_pf_1f_kg_2g_3),
\end{eqnarray}
where we introduced two auxiliary integrations over $l$ and $q$. The kinematics is shown in \fig{f2}, and
the expressions for $|M_{1p\leftrightarrow23k}^a|^2$ and $|M_{23\leftrightarrow1pk}^b|^2$ are given by
\begin{eqnarray}
|M^a_{1p \leftrightarrow 23k}|^2&=&64g^6N_c^3N_g\frac{(p\cdot p_1)^3}{q^2(q-k)^2(p\cdot k)(p_1\cdot k)},\non
|M^b_{23 \leftrightarrow 1kp}|^2&=&64g^6N_c^3N_g\frac{(p_2\cdot p_3)^3}{q^2(q+k)^2
(p_2\cdot k)(p_3\cdot k)}.
\end{eqnarray}

First, it is easy to show that, under the small angle approximation,
\begin{eqnarray}
q_0&\simeq&\bq\cdot\bv_1-\frac{{\bq'}_\perp^2}{2E_1}\non
&\simeq&E_k+(\bq-\bk)\cdot\bv_p+\frac{(\bq_\perp-\bk_\perp)^2}{2E_p},\\
l_0&\simeq&E_l+E_k-\bk\cdot\bv_l+\frac{{\bk''}_\perp^2}{2E_l}\non
&\simeq&E_p+q_0,\\
\bv_l&\simeq&\bv_p+\frac{\bq_\perp}{E_p}-\frac{\bq_\perp^2}{2E_p^2}\bv_p
+\frac{(\bq\cdot\bv_p)^2}{2E_p^2}\bv_p-\frac{\bq\cdot\bv_p}{E_p^2}\bq_\perp,
\end{eqnarray}
where $\bq_\perp=\bq-\bq\cdot\bv_p\bv_p$, $\bk_\perp=\bk-\bk\cdot\bv_p\bv_p$, $\bq'_\perp=\bq-\bq\cdot\bv_1\bv_1$, and ${\bk''}_\perp=\bk-\bk\cdot\bv_l\bv_l$.

Second, if $|\bk|<|\bq|$, then,
under collinear approximation, $\bv_k$ is nearly parallel to either $\bv_p$ or $\bv_1$.
For $\bv_k\simeq\bv_1$, $q\cdot k=|\bk|(q_0-\bq\cdot\bv_k)\simeq|\bk|[q_0-\bq\cdot\bv_1+\bq\cdot(\bv_1-\bv_k)]\sim-|\bk|{\bq'}^2_\perp/E_1
+|\bk||\bq|\h_{1k}\ll q^2$; For $\bv_k\simeq\bv_p$, $q\cdot k=|\bk|(|\bk|-\bk\cdot\bv_l)\sim|\bk|^2\h_{kp}^2+|\bk|\bq_\perp^2/(2E_p)+|\bk||\bq|\h_{kp}\ll q^2$. Thus in the small angle approximation plus the collinear approximation, if $|\bk|<|\bq|$, we can approximate $|M^a_{1p\leftrightarrow23k}|^2$ as
\begin{eqnarray}
|M^a_{1p\leftrightarrow23k}|^2&=&64g^6N_c^3N_g\frac{(p\cdot p_1)^3}{(q^2)^2(p\cdot k)(p_1\cdot k)}.
\end{eqnarray}
For $|M^b_{23\leftrightarrow1pk}|^2$, up to $q^2$ order, we can neglect $q\cdot k$ in the denominator and $k$ as well as $q\cdot p$
and $q\cdot p_1$ in the numerator,
\begin{eqnarray}
\label{mamb}
|M^b_{23\leftrightarrow1kp}|^2&=&64g^6N_c^3N_g\frac{[(p_1+q)\cdot (p+k-q)]^3}{q^2(q+k)^2
[(p+k-q)\cdot k][(p_1+q)\cdot k]}\non&\approx&
64g^6N_c^3N_g\frac{[(p_1+q)\cdot (p-q)]^3}{(q^2)^2
[(p-q)\cdot k][(p_1+q)\cdot k]}\non&\approx&
64g^6N_c^3N_g\frac{(p_1\cdot p)^3}{(q^2)^2
(p\cdot k)(p_1\cdot k)}\non&=&
|M^a_{1p\leftrightarrow23k}|^2.
\end{eqnarray}

Third, if $|\bk|>|\bq|$, under the collinear approximation, for $\bv_k\simeq\bv_1$, $q\cdot k\sim-|\bk|{\bq'}^2_\perp/E_1
+|\bk||\bq|\h_{1k}\ll q^2$ because $\h_{1k}\ll|{\bq'}|_\perp/E_1$; For $\bv_k\simeq\bv_p$, $q\cdot k\sim|\bk|^2\h_{kp}^2+|\bk|\bq_\perp^2/(2E_p)+|\bk||\bq|\h_{kp}\ll q^2$ because $\h_{kp}\ll|\bp_\perp|/E_p$. Thus the collinear approximation simplifies the matrix element
also when $|\bk|>|\bq|$:
\begin{eqnarray}
|M^a_{1p\leftrightarrow23k}|^2&=&64g^6N_c^3N_g\frac{(p\cdot p_1)^3}{(q^2)^2(p\cdot k)(p_1\cdot k)}\non
|M^b_{23\leftrightarrow1kp}|^2&=&64g^6N_c^3N_g\frac{[(p+k)\cdot p_1]^3}{(q^2)^2
(p\cdot k)(p_1\cdot k)}.
\end{eqnarray}

Fourth, the whole kinematic (phase) space can
be separated into two parts, one for $|\bk|<|\bq|$ and another for $|\bq|<|\bk|$.
We would expect that at kinematic region with $|\bk|<|\bq|$, the $2\leftrightarrow3$ process may be regarded as a $2\leftrightarrow2$ ``hard" process with one additional ``soft" gluon
emitted or absorbed by one of the ``hard" gluons; with $|\bq|<|\bk|$ the $2\leftrightarrow3$ process can be regarded as
an effective $1\leftrightarrow2$ process with a spectator gluon joined to make the effective $1\leftrightarrow2$ matrix element
nonzero (the matrix element of the $1\leftrightarrow2$ process is zero
for massless gluons). Thus we separate the collision kernel as
\begin{eqnarray}
{\cal C}_{2\leftrightarrow3}&=&{\cal C}^a_{2\leftrightarrow3}+{\cal C}^b_{2\leftrightarrow3}={\cal C}^>_{2\leftrightarrow3}+{\cal C}^<_{2\leftrightarrow3},
\end{eqnarray}
with
\begin{eqnarray}
{\cal C}^>_{2\leftrightarrow3}&=&\frac{1}{N_g}\int\frac{d^4q}{(2\p)^4}\int_{123l}\int_k^{q>k}\frac{2E_l}{2E_p}\int\frac{dl_0}
{2\p}(2\p)^8\d^4(p+q-l)\d^4(p_1-q-p_2)\non&&\times|M_{1p\leftrightarrow23k}^a|^2
[(2\p)^4\d^4(l-k-p_3)(g_pg_1f_2f_3f_k-f_pf_1g_2g_3g_k)\non&&
\;\;+(2\p)^4\d^4(l+k-p_3)(g_pg_1g_kf_2f_3-f_pf_1f_kg_2g_3)],\non
{\cal C}^<_{2\leftrightarrow3}&=&\frac{1}{N_g}\int\frac{d^4q}{(2\p)^4}
\int_{123l}\int_k^{q<k}\frac{2E_l}{2E_p}\int\frac{dl_0}
{2\p}(2\p)^8\d^4(p+q-l)\d^4(p_1-q-p_2)\non&&\times
[(2\p)^4\d^4(l-k-p_3)|M_{1p\leftrightarrow23k}^a|^2(g_pg_1f_2f_3f_k-f_pf_1g_2g_3g_k)\non&&
+(2\p)^4\d^4(l+k-p_3)|M_{23\leftrightarrow1kp}^b|^2(g_pg_1g_kf_2f_3-f_pf_1f_kg_2g_3)].\nonumber
\end{eqnarray}

\subsection {Simplifying ${\cal C}^>_{2\leftrightarrow3}$}
\label{c1}
Expand the integrand of ${\cal C}^>_{2\leftrightarrow3}$ in terms of $k$ and keep the leading order terms:
\begin{eqnarray}
{\cal C}^>_{2\leftrightarrow3}
&\approx&\frac{1}{N_g}\int\frac{d^4q}{(2\p)^4}\int_{12l}\int_k^{q>k}\frac{1}{2E_p}\int\frac{dl_0}
{2\p}(2\p)^8\d^4(p+q-l)\d^4(p_1-q-p_2)\non&&\times|M_{1p\leftrightarrow23k}^a|^2
[(2\p)\d(l_0-|\bk|-|\bl-\bk|)(g_pg_1f_2f_lf_k-f_pf_1g_2g_lg_k)
\non&&+(2\p)\d(l_0+|\bk|-|\bl+\bk|)(g_pg_1g_kf_2f_l-f_pf_1f_kg_2g_l)]\non
&\approx&\frac{1}{N_g}\int\frac{d^4q}{(2\p)^4}\int_{12l}\int_k^{q>k}\frac{1}{2E_p}\int\frac{dl_0}
{2\p}(2\p)^8\d^4(p+q-l)\d^4(p_1-q-p_2)\non&&\times|M_{1p\leftrightarrow23k}^a|^2
[(2\p)\d(l_0-|\bl|)(g_pg_1f_2f_lf_k-f_pf_1g_2g_lg_k)\non&&
+(2\p)\d(l_0-|\bl|)(g_pg_1g_kf_2f_l-f_pf_1f_kg_2g_l)]\non
&=&\frac{1}{N_g}\int\frac{d^4q}{(2\p)^4}\int_{12l}\int_k^{q>k}\frac{1}{2E_p}
(2\p)^4\d^4(p+q-l)(2\p)^4\d^4(p_1-q-p_2)\non&&\times|M_{1p\leftrightarrow2l}|^2
\frac{2g^2N_c(p\cdot p_1)}{(p\cdot k)(p_1\cdot k)}
(1+2f_k)(g_pg_1f_2f_l-f_pf_1g_2g_l),
\end{eqnarray}
where in the last equality $l^\m=(|\bl|,\bl)$ is on-shell, and the $2\leftrightarrow2$ matrix element is
\begin{eqnarray}
|M_{1p\leftrightarrow2l}|^2&=&32g^4N_c^2N_g\frac{(p\cdot p_1)^2}{(q^2)^2}.
\end{eqnarray}
We have written the $2\leftrightarrow3$ matrix element in a form of a $2\leftrightarrow2$ matrix element times a
$1\leftrightarrow2$ splitting function. Indeed,
if for example $k$ is nearly collinear to $p$,
\begin{eqnarray}
\frac{2g^2N_c(p\cdot p_1)}{(p\cdot k)(p_1\cdot k)}&\approx&\frac{2g^2N_c|\bp|}{(p\cdot k)|\bk|}\non
&=&\frac{2g^2}{(p+k)^2}P_{gg}(z)
\end{eqnarray}
with $P_{gg}(z)=2C_A/z$ being the standard unregularized $g\ra gg$ splitting function at $z\ra0$ limit where $z=E_k/E_p$~\cite{Altarelli:1977zs,Gribov:1972ri,Dokshitzer:1977sg}.

In the collinear approximation, if $\bv_\bk\simeq\bv_1$, then
$p\cdot p_1/k\cdot p\approx |\bp_1|/|\bk|$; or if $\bv_\bk\simeq\bv_p$, then
$p\cdot p_1/k\cdot p_1\approx |\bp|/|\bk|$. Thus we have
\begin{eqnarray}
\label{c>1}
{\cal C}^>_{2\leftrightarrow3}&\approx&\frac{1}{N_g}\int_{12l}\frac{1}{2E_p}|M_{1p\leftrightarrow2l}|^2
\int_{k<p_1-p_2}\frac{2g^2N_c}{|\bk|^2}
\ls\frac{1+2f_k}{1-\bv_k\cdot\bv_1}+\frac{1+2f_k}{1-\bv_k\cdot\bv_p}\rs\non
&&\times(2\p)^4\d^4(p+p_1-p_2-l)(g_pg_1f_2f_l-f_pf_1g_2g_l).
\end{eqnarray}
This is essentially a $p+p_1\leftrightarrow p_2+l$ collision kernel with an inner $1\leftrightarrow2$
splitting function. Let
\begin{eqnarray}
\label{dq}
{\cal D}(q)&=&\int_{k<q}\frac{2g^2N_c}{|\bk|^2}
\ls\frac{1+2f_k}{1-\bv_k\cdot\bv_1}+\frac{1+2f_k}{1-\bv_k\cdot\bv_p}\rs.
\end{eqnarray}
For isotropic distribution,
\begin{eqnarray}
\label{defD}
{\cal D}(q)
&=&2\int_{k<q}\frac{d^3\bk}{(2\p)^32E_k}\frac{2g^2N_c(1+2f_k)}{|\bk|^2(1-\cos\h)}\non
&=&2\frac{g^2N_c}{(2\p)^2}\int_0^{|\bq|} d |\bk|\frac{1+2f_k}{|\bk|}\int_0^\p d\h\frac{\sin\h}{1-\cos\h}\non
&=&2\frac{g^2N_c}{(2\p)^2}\int_0^{|\bq|} d |\bk|\frac{1+2f_k}{|\bk|}\int_{-1}^1 \frac{dx}{1-x}.
\end{eqnarray}
Thus,
\begin{eqnarray}
{\cal C}^>_{2\leftrightarrow3}&\approx&\frac{1}{N_g}\int_{12l}\frac{1}{2E_p}|M_{1p\leftrightarrow2l}|^2
{\cal D}(p_1-p_2)(2\p)^4\d^4(p+p_1-p_2-l)\non&&\times(g_pg_1f_2f_l-f_pf_1g_2g_l).
\end{eqnarray}
There are two types of infrared divergence in ${\cal D}(q)$. (1) The logarithmic divergence $\int_{-1}^1 dx/(1-x)\sim\int d\h/\h\sim\ln(1/\h_m)$ with $\h_m$ the minimal angle between $\bk$ and $\bp$. $\h_m$ arises completely due to interaction, so $\h_m\sim g$.
Thus, $\int_{-1}^1 dx/(1-x)\sim\ln(1/g)$ in both Glasma and nearly thermal equilibrium state. (2) Near thermal equilibrium, $f_k\sim T/\o_k$, thus
$\int^q_0 (dk/k)(1+2f_k)\sim2 T\int^q_0(dkk/(k^2+m_\infty^2)^{3/2}\sim T(1/m_\infty-1/m_D)$ where we use $m_\infty$ to denote the mass of the emitted or absorbed ultrasoft
gluon $k$ and $m_D$ to denote the mass of the exchanged gluon $q$. Near equilibrium, both $m_\infty$
and $m_D$ are of order $gT$ but can have different prefactors, we find
${\cal D}(q)\sim g\ln(1/g)$. In the initial Glasma, $m_D\sim m_\infty\sim Q_s$ and $f_k\sim 1/\a_s$, thus
$\int^q_0 (dk/k)(1+2f_k)\sim(2/\a_s)\ln(q/m_\infty)\sim 1/\a_s$. Thus ${\cal D}(q)\sim \ln(1/g)$. As the Glasma evolves, $f_k\sim \L_S/(\a_sk)$,
if $m_D\sim m_\infty\sim\sqrt{\L\L_S}$, thus $\int^q_0 (dk/k)(1+2f_k)\sim(1/\a_s)(\L_S/m_\infty)\sim(1/\a_s)\sqrt{\L_S/\L}$. Thus,
we find ${\cal D}(q)\sim \sqrt{\L_S/\L}\ln(1/g)$.

In either Glasma or nearly thermal equilibrium cases, we can conclude that the ratio of ultrasoft gluon emission and absorbtion $2\leftrightarrow3$ processes over the purely elastic $2\leftrightarrow2$ processes is either $\ln(1/g)$ order or $g\ln(1/g)$ order.

\subsection {Simplifying ${\cal C}^<_{2\leftrightarrow3}$}
\label{c2}
Expand the distribution functions in ${\cal C}^<_{2\leftrightarrow3}$ in terms of $q$ and keep the leading order terms:
\begin{eqnarray}
{\cal C}^<_{2\leftrightarrow3}
&\approx&\frac{1}{N_g}\int\frac{d^4q}{(2\p)^4}
\int_{123}\int_k^{q<k}\frac{1}{2E_p}(2\p)^4\d^4(p_1-q-p_2)h_1\non&&\times
[(2\p)^4\d^4(p+q-k-p_3)|M_{1p\leftrightarrow23k}^a|^2(g_pf_3f_k-f_pg_3g_k)\non&&
+(2\p)^4\d^4(p+q+k-p_3)|M_{23\leftrightarrow1kp}^b|^2(g_pg_kf_3-f_pf_kg_3)]\non
&\approx&\frac{1}{N_g}\int\frac{d^4q}{(2\p)^4}
\int_{13}\int_k^{q<k}\frac{1}{2E_p2E_1}(2\p)
\d(q_0-\bq\cdot\bv_1)h_1\non&&\times
[(2\p)^4\d^4(p+q-k-p_3)|M_{1p\leftrightarrow23k}^a|^2(g_pf_3f_k-f_pg_3g_k)\non&&
+(2\p)^4\d^4(p+q+k-p_3)|M_{23\leftrightarrow1kp}^b|^2(g_pg_kf_3-f_pf_kg_3)],\nonumber
\end{eqnarray}
where $h_1\equiv f_1 g_1$. This is basically a $1\leftrightarrow2$ collision kernel with the $2\leftrightarrow2$ processes playing a role of opening a finite phase space for $1\leftrightarrow2$ process. Because $q$ is small, $p, k$ and
$p_3$ are nearly collinear (there is vanishing
phase space at the collinear region $k\parallel p_1$). In
the collinear approximation, we have
\begin{eqnarray}
p_0+q_0-k_0-p_{30}&\approx&q_0-|\bk|-(\bq-\bk)\cdot\bv_p\approx q_0-\bq\cdot\bv_p,\non
p_0+q_0+k_0-p_{30}&\approx&q_0+|\bk|-(\bq+\bk)\cdot\bv_p\approx q_0-\bq\cdot\bv_p.\nonumber
\end{eqnarray}
The matrix element are then
\begin{eqnarray}
|M^a_{1p\leftrightarrow23k}|^2&=&64g^6N_c^3N_g\frac{(p\cdot p_1)^2}{(q^2)^2|\bk|^2(1-\bv_p\cdot\bv_k)},\\
|M^b_{23\leftrightarrow1kp}|^2&=&64g^6N_c^3N_g\frac{|\bp|+|\bk|}{|\bk|}\frac{(p\cdot p_1)^2}{(q^2)^2|\bk|^2(1-\bv_p\cdot\bv_k)}.
\end{eqnarray}
Thus,
\begin{eqnarray}
{\cal C}^<_{2\leftrightarrow3}
\label{c<1}
&\approx&\frac{1}{N_g}\int\frac{d^4q}{(2\p)^4}
\int_{1}\int_k^{q<k}\frac{1}{(2E_p)^22E_1}(2\p)
\d(q_0-\bq\cdot\bv_1)(2\p)
\d(q_0-\bq\cdot\bv_p)h_1\non&\times&
\ls|M_{1p\leftrightarrow23k}^a|^2(g_pf_{p-k}f_k-f_pg_{p-k}g_k)
+|M_{23\leftrightarrow1kp}^b|^2(g_pg_kf_{p+k}-f_pf_kg_{p+k})\rs.\non
\end{eqnarray}
In the following we denote $q=|\bq|, p=|\bp|, p_1=|\bp_1|, k=|\bk|$. Let $q_0=x q$ and let $\bv_p=(1,0,0)$, $\bv_1=(\cos\h_1,\sin\h_1,0)$, and $\bv_q=(\sin\h_q\cos\f_q,
\sin\h_q\sin\f_q,\cos\h_q)$. We have
\begin{eqnarray}
\label{c<int}
&&\d(x-\bv_q\cdot\bv_1)\d(x-\bv_q\cdot\bv_p)=\d[x-\sin\h_q\cos(\h_1-\f_q)]\d(x-\sin\h_q\cos\f_q)\non
&=&\frac{1}{\sin\h_q}\frac{\d(x-\sin\h_q\cos\f_q)}{|\sin\f_q-\sin(\f_q-\h_1)|}
\ls\d\lb\f_q-\frac{\h_1}{2}\rb
+
\d\lb\f_q-\frac{\h_1}{2}-\p\rb\rs\non
&=&\frac{1}{\sin\h_q}\frac{1}{2\sin(\h_1/2)}
\bigg[\d\lb\f_q-\frac{\h_1}{2}\rb
\d\lb x-\sin\h_q\cos\frac{\h_1}{2}\rb\non&&+
\d\lb\f_q-\frac{\h_1}{2}-\p\rb
\d\lb x+\sin\h_q\cos\frac{\h_1}{2}\rb\bigg].\nonumber
\end{eqnarray}
Thus,
\begin{eqnarray}
&&\int\frac{dq_0}{2\p}d\O_q(2\p)
\d(q_0-\bq\cdot\bv_1)(2\p)
\d(q_0-\bq\cdot\bv_p)|M_{1p\ra23k}^a|^2\non&=&128\p g^6N_c^3N_g\frac{(pp_1)^2}{q^5k^2}
\int dx\int_0^\p d\h_q\sin\h_q\int_0^{2\p}d\f_q
\d(x-\bv_q\cdot\bv_1)\d(x-\bv_q\cdot\bv_p)\non&&\times\frac{(
1-\bv_p\cdot\bv_1)^2}{(1-x^2)^2(1-\bv_p\cdot\bv_k)}\non&=&
128\p g^6N_c^3N_g\frac{(pp_1)^2}{q^5k^2}\frac{(1-\bv_p\cdot\bv_1)^2}{1-\bv_p\cdot\bv_k}
\int_0^\p d\h_q
\frac{1}{\sin\frac{\h_1}{2}(
1-\sin^2\h_q\cos^2\frac{\h_1}{2})^2}\non
&=&
128\p^2 g^6N_c^3N_g\frac{(pp_1)^2}{q^5k^2}
\frac{3-\bv_p\cdot\bv_1}{1-\bv_p\cdot\bv_k}.
\end{eqnarray}
Furthermore, for isotropic distributions, we have:
\begin{eqnarray}
{\cal C}^<_{2\leftrightarrow3}
&=&\frac{128\p^2 g^6N_c^3N_g}{N_g}\int_0^\infty\frac{dq q^2}{(2\p)^3}
\int_{1}\int_k^{q<k}\frac{1}{(2E_p)^22E_1}\frac{(pp_1)^2}{q^5k^2}
\frac{3-\bv_p\cdot\bv_1}{1-\bv_p\cdot\bv_k}h_1\non&&\times
\ls(g_pf_{p-k}f_k-f_pg_{p-k}g_k)
+\frac{(p+k)^3}{p^3}(g_pg_kf_{p+k}-f_pf_kg_{p+k})\rs\non
&=&\frac{3g^6N_c^3}{16\p^5}
\int_0^\infty dp_1 p_1^2h_1\int_{-1}^1\frac{dx}{1-x}\int_0^\infty \frac{dq}{q^3}\int_{q}^\infty \frac{dk}{k}
\non&&\times
\ls(g_pf_{p-k}f_k-f_pg_{p-k}g_k)
+\frac{(p+k)^3}{p^3}(g_pg_kf_{p+k}-f_pf_kg_{p+k})\rs,
\end{eqnarray}
where the upper limit of the integration over $k$ for the first two terms should be cut at $p$.
When $k$ is small the integrand over $k$ goes like $(1+2f_p)f'_p$. It is finite, so we can put the
lower limit of $\int dk$ as $0$. In the first two terms, let $k=zp$ with $z$ being the momentum fraction of the emitted gluon; in the
last two terms let $k=z(p+k)$ with $z$ being the momentum fraction of the absorbed gluon. Then we have
\begin{eqnarray}
{\cal C}^<_{2\leftrightarrow3}
&=&\x\a_s^2 R\frac{I_a}{I_b}\int_{0}^{z_c} \frac{dz}{z}\bigg\{
\ls g_pf_{(1-z)p}f_{zp}-f_pg_{(1-z)p}g_{zp}
\rs\non&&+
\frac{1}{(1-z)^4}\ls g_pg_{zp/(1-z)}f_{p/(1-z)}-f_pf_{zp/(1-z)}g_{p/(1-z)})\rs\bigg\},
\end{eqnarray}
where we introduce the momentum fraction cut $z_c<1$ to characterizing the fact that $k$ is a small fraction
of the total momentum in this effective $1\leftrightarrow2$ process and the prefactor $R$ is given by
\begin{eqnarray}
\label{defR}
R\equiv\frac{12N_c^3}{\p^2}
\frac{1}{\x}\int_{-1}^1\frac{dx}{1-x}m_D^2\int_0^\infty \frac{dq}{q^3}\sim\hat{O}(1).
\end{eqnarray}
There are two kinds of infrared divergences in ${\cal C}^<_{2\leftrightarrow3}$:
(1) The logarithmic divergence: $\int_{-1}^1 dx/(1-x)\sim\ln(1/g)$. (2) The quadratic divergence $\int dq/q^3\sim1/m_D^2$. Noticing that ${\cal C}_{2\leftrightarrow2}$ is of order $\a_s^2\ln(1/g)$
near equilibrium and $(Q_s/\a_s)\ln(1/g)$ in initial Glasma state, we
find that ${\cal C}^<_{2\leftrightarrow3}/{\cal C}_{2\leftrightarrow2}$ is of order $\hat{O}(1)$ in both equilibrium and initial Glasma states. It is worth  mentioning that recent analysis in \cite{Zhang:2012vi} has shown that the the $2\leftrightarrow3$ inelastic cross section from exact matrix element becomes significantly smaller than that from the Gunion-Bertsch formula, and amounts to $\sim 20\%$ of the $2\leftrightarrow2$ cross section. It therefore seems very plausible that a realistic choice of $R$ value shall be rather modest.

Now we show that the collision kernel ${\cal C}^<_{2\leftrightarrow3}$ conserves energy, i.e,\\
$\int_0^\infty dp p^3{\cal C}^<_{2\leftrightarrow3}[f_p]=0$:
\begin{eqnarray}
\int_0^\infty dp p^3{\cal C}^<_{2\leftrightarrow3}[f_p]
&\propto&\int_0^\infty dp p^3
\int_{0}^{z_c} \frac{dz}{z}\bigg\{
\ls g_pf_{(1-z)p}f_{zp}-f_pg_{(1-z)p}g_{zp}
\rs\non&&+
\frac{1}{(1-z)^4}\ls g_pg_{zp/(1-z)}f_{p/(1-z)}-f_pf_{zp/(1-z)}g_{p/(1-z)})\rs\bigg\}.\nonumber
\end{eqnarray}
In the last two terms, by changing the variable $p\ra (1-z)p$, one finds that the first two terms
cancel the last two terms so that $\int_0^\infty dp p^3{\cal C}^<_{2\leftrightarrow3}[f_p]=0$.

Some remarks are in order regarding the effective reduction of ${\cal C}^>_{2\leftrightarrow3}$ to an essentially elastic contribution.  It shall be noted that the whole kernel ${\cal C}_{2\leftrightarrow3}$ certainly is and should be number changing overall. However leading contributions from certain specific kinetic domain may not necessarily be so.   What we have shown is  that under the small-angle and collinear approximation the $2\leftrightarrow3$ collision kernel ${\cal C}_{2\leftrightarrow3}$ can be split into two parts in correspondence to two different kinematic domains, ${\cal C}_{2\leftrightarrow3}={\cal C}^<_{2\leftrightarrow3}+{\cal C}^>_{2\leftrightarrow3}$, where
${\cal C}^<_{2\leftrightarrow3}$ is an effective $1\leftrightarrow2$ kernel and ${\cal C}^>_{2\leftrightarrow3}$ becomes effectively   elastic. In such a way, we encode the dominant inelastic effects into ${\cal C}^<_{2\leftrightarrow3}$ and the role of ${\cal C}^>_{2\leftrightarrow3}$ is to renormalize the total rate of the $2\leftrightarrow2$ process. So  why the piece of contribution ${\cal C}^>_{2\leftrightarrow3}$ that originally emerges from the  inelastic kernel ${\cal C}_{2\leftrightarrow3}$ becomes effectively elastic? This is because in the kinematic region for ultrasoft gluon emission and absorption, $|\bk|\ll|\bq|$, the matrix element for \fig{f2} (left), $|M^a_{1p\leftrightarrow23k}|^2$, is equal to that of \fig{f2} (right), $|M^b_{23\leftrightarrow1kp}|^2$, see \eq{mamb}. Intuitively this may be understood as follows: on top of a 2 to 2 scattering, one may attach an extremely soft particle either on one incoming particle (thus making a $3\to 2$ contribution) or on one outgoing particle (thus making a $2\to 3$ contribution), but the two processes have the same rate  and thus cancel out to the leading order of $|\bk|$. If one includes even higher orders of the expansion in terms of $|\bk|$ there would be sub-leading number-changing contributions from ${\cal C}^>_{2\leftrightarrow3}$ as well. To the leading order of small-angle and collinear approximation that we consider here, there is clearly advantage in doing such a  careful separation of contributions from different regions of the phase space.

\section {The kinetic equation for anisotropic system}
\label{aniso}
\subsection {Simplify ${\cal C}^>_{2\leftrightarrow3}$ for anisotropic system}
\label{c1aniso}
Although in this paper we mainly focus on the isotropic system, we will in this Appendix present the kinetic equation for anisotropic system.
In the anisotropic case, \eqs{c>1}{dq} are still valid. (When there is no confusion, we will use $k$ to denote
$|\bk|$ and also the four momentum $k$.
Somewhere, we will use $f_k$ to denote $f(\bk)$.)
\begin{eqnarray}
{\cal C}^>_{2\leftrightarrow3}&\approx&\frac{1}{N_g}\int_{12l}\frac{1}{2E_p}
\int\frac{d^4q}{(2\p)^4}|M_{1p\leftrightarrow2l}|^2{\cal D}(q)(2\p)^4\d^4(p_1-q-p_2)\non
&&\times(2\p)^4\d^4(p+q-l)(g_pg_1f_2f_l-f_pf_1g_2g_l).
\end{eqnarray}
\begin{eqnarray}
{\cal D}(q)&=&\int_{k<q}\frac{2g^2N_c}{|\bk|^2}
\ls\frac{1+2f_k}{1-\bv_k\cdot\bv_1}+\frac{1+2f_k}{1-\bv_k\cdot\bv_p}\rs\non
&\approx&\int_{k<q}\frac{d^3\bk}{(2\p)^32E_k}\frac{2g^2N_c}{|\bk|^2}
\ls\frac{1+2f(k\bv_1)}{1-\bv_k\cdot\bv_1}+\frac{1+2f(k\bv_p)}{1-\bv_k\cdot\bv_p}\rs\non
&=&\frac{2g^2N_c}{(2\p)^2}\int_0^{q} d k\frac{1+f(k\bv_1)+f(k\bv_p)}{k}\int_{-1}^1 \frac{dx}{1-x}.
\end{eqnarray}
First, we show that ${\cal C}^>_{2\leftrightarrow3}$ conserves particle number. To see this, we write
\begin{eqnarray}
{\cal D}(q)&=&{\cal D}_1(q)+{\cal D}_2(q),\non
{\cal D}_1(q)
&=&\frac{2g^2N_c}{(2\p)^2}\int_0^{q} d k\frac{2+f(k\bv_1)+f(k\bv_p)+f(k\bv_2)+f(k\bv_l)}{2k}\int_{-1}^1 \frac{dx}{1-x},\non\\
{\cal D}_2(q)
&=&\frac{2g^2N_c}{(2\p)^2}\int_0^{q} d k\frac{f(k\bv_1)+f(k\bv_p)-f(k\bv_2)-f(k\bv_l)}{2k}\int_{-1}^1 \frac{dx}{1-x}.\non
\end{eqnarray}
Expand $\bv_2$ around $\bv_1$ and $\bv_l$ around $\bv_p$:
\begin{eqnarray}
\bv_2&\approx&\bv_1-\frac{\bq-\bq\cdot\bv_1\bv_1}{p_1}-\frac{[q^2-3(\bq\cdot\bv_1)^2]\bv_1+\bv_1\cdot\bq\bq}
{2p_1^2},\\
\bv_l&\approx&\bv_p+\frac{\bq-\bq\cdot\bv_p\bv_p}{p}-\frac{[q^2-3(\bq\cdot\bv_p)^2]\bv_p+\bv_p\cdot\bq\bq}
{2p^2}.
\end{eqnarray}
Thus
\begin{eqnarray}
f(k\bv_1)-f(k\bv_2)&=&k(\bv_1-\bv_2)\cdot\bv_1\frac{\pt}{\pt k}f(k\bv_1)\non&&+\frac{k^2}{2}[(\bv_1-\bv_2)\cdot\bv_1]^2\frac{\pt^2}{\pt k^2}f(k\bv_1)+\cdots\non
&=&\frac{q^2-2(\bq\cdot\bv_1)^2}{2p^2_1}k\frac{\pt}{\pt k} f(k\bv_1) +O\lb\frac{q}{p_1}\rb^3,\\
f(k\bv_p)-f(k\bv_l)
&=&\frac{q^2-2(\bq\cdot\bv_p)^2}{2p^2}k\frac{\pt}{\pt k} f(k\bv_p) +O\lb\frac{q}{p}\rb^3,
\end{eqnarray}
Because $q/p_1$, $q/p$ are small, we have
\begin{eqnarray}
{\cal D}(q)
&\approx&{\cal D}_1(q).
\end{eqnarray}
Then we have
\begin{eqnarray}
&&\int\frac{d^3\bp}{(2\p)^3}{\cal C}^>_{2\leftrightarrow3}[f_p]\non
&\approx&\frac{1}{N_g}\int_{12pl}
\int\frac{d^4q}{(2\p)^4}|M_{1p\leftrightarrow2l}|^2{\cal D}_1(q)(2\p)^4\d^4(p_1-q-p_2)\non
&&\times(2\p)^4\d^4(p+q-l)(g_pg_1f_2f_l-f_pf_1g_2g_l)\non
&=&\frac{H}{N_g}\int_{12pl}
\int\frac{d^4q}{(2\p)^4}|M_{1p\leftrightarrow2l}|^2\int_0^{q} d k\frac{2+f(k\bv_1)+f(k\bv_p)+f(k\bv_2)+f(k\bv_l)}{2k}\non
&&\times(2\p)^4(2\p)^4\d^4(p_1-q-p_2)\d^4(p+q-l)(g_pg_1f_2f_l-f_pf_1g_2g_l),
\end{eqnarray}
with
\begin{eqnarray}
H=\frac{2g^2N_c}{(2\p)^2}\int_{-1}^1 \frac{dx}{1-x}.
\end{eqnarray}
Because the integrand is anti-symmetric in $(1,p)$ and $(2,l)$, so the integral vanishes.

Thus, ${\cal C}^>_{2\ra3}$ can be written in a form $-\nabla_p\cdot {\bf S}$ where the flux $S^i$ is
\begin{eqnarray}
S^i=\frac{N_c^2}{4\p}g^4H\int\frac{dq}{q}\int\frac{d^3\bp_1}{(2\p)^3}
\int_0^{q} d k\frac{1+f(k\bv_1)+f(k\bv_p)}{k}\ls h_p\nabla_1^j f_{1}-h_{1}\nabla_p^j f_p
\rs{\cal V}^{ij},\non
\end{eqnarray}
with the tensor
\begin{eqnarray}
{\cal V}^{ij}=\d^{ij}(1-\bv_p\cdot\bv_1)+\lb v_p^i v_1^j+v_p^j v_1^i\rb.
\end{eqnarray}
There are now two terms in $S^i$. The second term can be simplified as
\begin{eqnarray}
\sim-\frac{N_c^2}{4\p}g^4H\nabla_p^i f_p\int\frac{dq}{q}\int\frac{d^3\bp_1}{(2\p)^3}
\int_0^{q} d k\frac{1+f(k\bv_1)+f(k\bv_p)}{k} h_{1},
\end{eqnarray}
where we have used the property $f(\bp)=f(-\bp)$ to cancel all terms linear in $\bv_1$.
For the first term, because $f(\bp)=f(-\bp)$, $\nabla_1^jf_1\propto p_1^j$, the only nonzero
contributions in ${\cal V}^{ij}\nabla_1^j$ should be
\begin{eqnarray}
{\cal V}^{ij}\nabla_1^j&\sim&- v_p^j v_1^j\nabla_1^i+(v_p^i v_1^j+v_p^j v_1^i)\nabla_1^j\non
&\sim&- v_p^i v_1^i\nabla_1^i+v_p^i v_1^j\nabla_1^j+v_p^i v_1^i\nabla_1^i\non
&\sim& v_p^i v_1^j\nabla_1^j=v_p^i\bv_1\cdot\nabla_1=v_p^i\frac{\pt}{\pt p_1}.
\end{eqnarray}
Then the first term is
\begin{eqnarray}
&\sim&\frac{N_c^2}{4\p}g^4Hh_p v_p^i\int\frac{dq}{q}\int\frac{d^3\bp_1}{(2\p)^3}
\int_0^{q} \frac{d k}{k}\frac{\pt f_1}{\pt p_1}[1+f(k\bv_1)+f(k\bv_p)]\non
&=&-\frac{N_c^2}{4\p}2g^4Hh_pv_p^i\int\frac{dq}{q}\int_0^{q} \frac{d k}{k}\int\frac{d^3\bp_1}{(2\p)^3}
f_{1}\frac{1+f(k\bv_1)+f(k\bv_p)}{p_1},
\end{eqnarray}
where to arrive at the second line we have made integration by part over $p_1$.
The flux is then
\begin{eqnarray}
S^i&=&\frac{N_c^2}{4\p}g^4H\bigg\{-2h_pv_p^i\int\frac{dq}{q}\int_0^{q} \frac{d k}{k}\int\frac{d^3\bp_1}{(2\p)^3}
f_{1}\frac{1+f(k\bv_1)+f(k\bv_p)}{p_1}\non&&-\nabla_p^i f_p\int\frac{dq}{q}\int\frac{d^3\bp_1}{(2\p)^3}
\int_0^{q} d k\frac{1+f(k\bv_1)+f(k\bv_p)}{k} h_{1}\bigg\}\non
&\approx&\frac{N_c^2}{4\p}g^4H\bigg\{-2h_pv_p^i\int\frac{dq}{q}\int_0^{m_D} \frac{d k}{k}\int\frac{d^3\bp_1}{(2\p)^3}
f_{1}\frac{1+f(k\bv_1)+f(k\bv_p)}{p_1}\non&&-\nabla_p^i f_p\int\frac{dq}{q}\int\frac{d^3\bp_1}{(2\p)^3}
\int_0^{m_D} d k\frac{1+f(k\bv_1)+f(k\bv_p)}{k} h_{1}\bigg\}\non
&=&\frac{N_c^2}{4\p}g^4LH\bigg\{-2h_pv_p^i\int\frac{d^3\bp_1}{(2\p)^3}
f_{1}\frac{K(\bv_1,\bv_p)}{p_1}-\nabla_p^i f_p\int\frac{d^3\bp_1}{(2\p)^3}
K(\bv_1,\bv_p) h_{1}\bigg\},\non
\end{eqnarray}
where we define
\begin{eqnarray}
L&\equiv&\int\frac{dq}{q},\\
K(\bv_1,\bv_p)&\equiv&\int_0^{m_D} d k\frac{1+f(k\bv_1)+f(k\bv_p)}{k}.
\end{eqnarray}
Thus ${\cal C}^>_{2\leftrightarrow3}$ becomes
\begin{eqnarray}
{\cal C}^>_{2\leftrightarrow3}&=&-\nabla_p\cdot {\bf S}\non
&=&\frac{N_c^2}{4\p}g^4LH\nabla_p\cdot\bigg\{2h_p\bv_p\int\frac{d^3\bp_1}{(2\p)^3}
f_{1}\frac{K(\bv_1,\bv_p)}{p_1}\non&&+(\nabla_p f_p)\int\frac{d^3\bp_1}{(2\p)^3}
K(\bv_1,\bv_p) h_{1}\bigg\}.
\end{eqnarray}
It obviously conserves particle number and it is not difficult to show that it conserves energy as well.
Furthermore, the Bose-Einstein distribution with an arbitrary chemical potential
vanishes ${\cal C}^>_{2\leftrightarrow3}$.

\subsection {Simplify ${\cal C}^<_{2\leftrightarrow3}$ for anisotropic system}
\label{c2iso}
In this case, \eqs{c<1}{c<int} are still valid, we have
\begin{eqnarray}
{\cal C}^<_{2\leftrightarrow3}
&=&\frac{2g^6N_c^3}{\p}\int_0^\infty dq
\int_{1}\int_k^{q<k}\frac{p_1}{q^3k^2}
\frac{3-\bv_p\cdot\bv_1}{1-\bv_p\cdot\bv_k}h_1\non&&\times
\bigg\{[g(\bp)f(\bp-k\bv_p)f(k\bv_p)-f(\bp)g(\bp-k\bv_p)g(k\bv_p)]\non
&&+\frac{(p+k)^3}{p^3}[g(\bp)g(k\bv_p)f(\bp+k\bv_p)-f(\bp)f(k\bv_p)g(\bp+k\bv_p)]\bigg\}\non
&=&\frac{g^6N_c^3}{2\p}\int_0^\infty dq
\int\frac{d^3\bp_1}{(2\p)^3}h_1\int_{q}^\infty dk\frac{d\O_k}{(2\p)^3}\frac{1}{q^3k}
\frac{3-\bv_p\cdot\bv_1}{1-\bv_p\cdot\bv_k}\non&&\times
\bigg\{[g(\bp)f(\bp-k\bv_p)f(k\bv_p)-f(\bp)g(\bp-k\bv_p)g(k\bv_p)]\non
&&+\frac{(p+k)^3}{p^3}[g(\bp)g(k\bv_p)f(\bp+k\bv_p)-f(\bp)f(k\bv_p)g(\bp+k\bv_p)]\bigg\}\non
&=&\frac{3g^6N_c^3}{2\p}\int_0^\infty dq
\int\frac{d^3\bp_1}{(2\p)^3}h_1\int_{q}^\infty dk\frac{d\O_k}{(2\p)^3}\frac{1}{q^3k}
\frac{1}{1-\bv_p\cdot\bv_k}\non&&\times
\bigg\{[g(\bp)f(\bp-k\bv_p)f(k\bv_p)-f(\bp)g(\bp-k\bv_p)g(k\bv_p)]\non
&&+\frac{(p+k)^3}{p^3}[g(\bp)g(k\bv_p)f(\bp+k\bv_p)-f(\bp)f(k\bv_p)g(\bp+k\bv_p)]\bigg\}\non
&=&\frac{3g^6N_c^3}{(2\p)^3}\int\frac{d^3\bp_1}{(2\p)^3}h_1
\int_{-1}^1\frac{dx}{1-x}\int_0^\infty \frac{dq}{q^3}\int_{q}^\infty \frac{dk}{k}\non&&\times
\bigg\{[g(\bp)f(\bp-k\bv_p)f(k\bv_p)-f(\bp)g(\bp-k\bv_p)g(k\bv_p)]\non
&&+\frac{(p+k)^3}{p^3}[g(\bp)g(k\bv_p)f(\bp+k\bv_p)-f(\bp)f(k\bv_p)g(\bp+k\bv_p)]\bigg\},\non
\end{eqnarray}
where the upper limit of the integration over $k$ for the first two terms should be cut at $p$.
The lower limit of the integration over $k$ can be set to be zero because there is no IR singularity.
In the first two terms, let $k=zp$ with $z$ being the momentum fraction of the emitted gluon; in the
last two terms let $p=(1-z)(p+k)$. Then we have
\begin{eqnarray}
{\cal C}^<_{2\leftrightarrow3}
&=&\frac{3g^6N_c^3}{(2\p)^3}
\int\frac{d^3\bp_1}{(2\p)^3}h_1\int_{-1}^1\frac{dx}{1-x}\int_0^\infty \frac{dq}{q^3}\non&&\times
\int_{0}^1 \frac{dz}{z(1-z)}\bigg\{\frac{1}{2}
\ls g_\bp f_{(1-z)\bp}f_{z\bp}-f_\bp g_{(1-z)\bp}g_{z\bp}
\rs\non&&+
\frac{1}{(1-z)^3}\ls g_\bp g_{z\bp/(1-z)}f_{\bp/(1-z)}-f_\bp f_{z\bp/(1-z)}g_{\bp/(1-z)})\rs\bigg\}.
\end{eqnarray}
It can be shown that the collision kernel ${\cal C}^<_{2\leftrightarrow3}$ conserves energy, i.e,
$\int d^3\bp p {\cal C}^<_{2\leftrightarrow3}[f_p]=0$.

\end{document}